\documentclass[11pt]{article}

\usepackage{graphicx}
\usepackage{caption}
\usepackage{subcaption}
\usepackage{lineno}
\usepackage{float}

\usepackage{verbatim}
\usepackage{amsmath}
\usepackage{amsfonts}
\usepackage{amssymb}
\usepackage{color}
\usepackage{hyperref}
 \usepackage{stackrel}
\usepackage{upgreek}
 \usepackage{csquotes}
 \usepackage{ulem}
 \usepackage{cite}

 \usepackage{fullpage}

 \def\##1{{\bf #1}}
\renewcommand\vec{\mathbf}

\def\=#1{\underline{\underline{#1}}}

\definecolor{Lak}{rgb}{0.7,0.3,0.7}

\def\muo{\mu_{\scriptscriptstyle 0}}

\def\etao{\eta_{\scriptscriptstyle 0}}

\def\mur{\mu_{\rm r}}

\def\ux{\hat{\#x}}
\def\uy{\hat{\#y}}
\def\uz{\hat{\#z}}

\def\um{\hat{\#m}}

\def\bro{{\#r}_{\rm o}}
\def\xio{\xi_{\rm o}}
\def\etao{\eta_{\rm o}}
\def\phio{\phi_{\rm o}}
\def\etao{\eta_{\rm o}}

\def\Ro{R_{\rm o}}
\def\ro{r_{\rm o}}
\def\Vminus{V^-}
\def\Vplus{V^+}

\def\.{\mbox{ \tiny{$^\bullet$} }}

\newenvironment{rcases}
  {\left.\begin{aligned}}
  {\end{aligned}\right\rbrace}

\def\les{\left[}
\def\ris{\right]}
\def\lec{\left\{}
\def\ric{\right\}}

\def\calA{{\cal A}}
\def\calB{{\cal B}}
\def\calC{{\cal C}}
\def\calD{{\cal D}}

\def\calI{{\cal I}}
\def\calJ{{\cal J}}
\def\calK{{\cal K}}
\def\calL{{\cal L}}
\def\calM{{\cal M}}
\def\calN{{\cal N}}
\def\calT{{\cal T}}

\def\calS{{\cal S}}
\def\calU{{\cal U}}

\def\inc{_{\rm source}}
\def\sca{_{\rm pert}}

\def\jk{ jk}
\def\mn{ {\ell}n}
\def\jkmn{_{jk{\ell}n}}

\def\Psione{\Psi_{\mn}^{1 \jk}}
\def\Psitwo{\Psi_{\mn}^{2 \jk}}
\def\Psionep{\Psi_{\ell^\prime n^\prime}^{1 j^\prime k^\prime}}
\def\uxi{\hat{\#\xi}}
\def\uxi{\hat{\boldsymbol{\upxi}}}
\def\double{_{ j j^\prime k k^\prime \ell\ell^\prime nn^\prime}}

\def\msx{m_{\rm sx}}
\def\msy{m_{\rm sy}}
\def\msz{m_{\rm sz}}

\begin{document}

\begin{center}
\textbf{Theory of perturbation of the magnetostatic field by an anisotropic  magnetic toroid} \\[10pt]
\textit{Hamad M. Alkhoori}\\
{Department of Electrical Engineering, United Arab Emirates University, P.O. Box 15551, Al Ain, UAE} \footnote{Corresponding author; e-mail: hamad.alkhoori@uaeu.ac.ae}\\
\textit{Akhlesh Lakhtakia}\\
{Department of Engineering Science and Mechanics, The Pennsylvania State University,
University Park, Pennsylvania 16802, USA}\\
 \textit{Nikolaos L. Tsitsas}\\
{School of Informatics, Aristotle University of Thessaloniki, Thessaloniki 54124, Greece}\\

\end{center}

\begin{abstract}
The perturbation of a magnetostatic field by a toroid made of a homogeneous anisotropic magnetic material was formulated using the solutions of the Laplace equation in the toroidal coordinate system. That was straightforward
in the region outside the toroid, but an affine coordinate transformation had to be employed inside the toroid. The
coefficients of the series expansion of the perturbation potential in terms of appropriate toroidal basis functions
were related to the coefficients of the series expansion of the source potential in terms of appropriate toroidal basis functions by a transition matrix. As a
result of the solution of this novel problem, the consequences of material anisotropy on perturbing the magnetostatic field are
clearly evident in the region near the toroid.

 \end{abstract}

\section{Introduction}

In a source-free region, the magnetostatic problem requires  solution of the
Gauss equation $\nabla \. \vec{B}(\vec{r}) = 0$ and the Ampere equation
$\nabla \times \vec{H}(\vec{r})= \vec{0}$, where $\vec{H}(\vec{r})$ is the magnetostatic field and $\vec{B}(\vec{r})$ is the magnetostatic flux density. The Ampere equation allows the definition of a scalar magnetic potential $\Phi(\#r)$ through
\begin{equation} \label{H}
\vec{H}(\#r) =  - \nabla \Phi(\#r)\,.
\end{equation}
Suppose that $\vec{B}(\vec{r}) =\mu \vec{H}(\vec{r})$ in a certain region filled with a homogeneous isotropic medium
with permeability scalar $\mu$. Then, the scalar magnetic potential satisfies the Laplace equation
\begin{equation}
\label{Laplace}
\nabla^2\Phi(\#r)=0
\end{equation}
in that region.

The Laplace equation can be solved by the method of separation of variables in thirteen three-dimensional coordinate systems: Cartesian, confocal paraboloidal, bispherical, toroidal, and
ten specializations of the confocal ellipsoidal system \cite{Eisenhart,Moon}. Accordingly, magnetostatic boundary-value problems can be solved, at least in principle, so long as the  permeability scalar is piecewise uniform with discontinuities only across boundaries on which one of the three coordinates is invariant.

What if the permeability is not  the scalar $\mu$ in a certain region filled with a homogeneous medium, but the dyadic $\=\mu$? In other words, the medium is anisotropic \cite{EAB,Chen}. Then we get
\begin{equation}
\nabla\.\left[\=\mu\.\nabla\Phi(\#r)\right]=0
\end{equation}
in that region, so that $\Phi(\#r)$ is not a solution of the Laplace equation in that region.
The solutions of Eq.~(\ref{Laplace}) can still be applied though an affine coordinate transformation.

For the sake of illustration, in this paper we have focused on the perturbation of a
magnetostatic field $\vec{H}_\text{source}$
by a toroid made of a homogeneous anisotropic material. Perturbation by anisotropic magnetic toroids
does not appear to have been treated heretofore, despite the fact that magnetic
 anisotropy is exhibited
by both crystalline materials \cite{Gersten,McCaig,EAB}
and effectively homogeneous composite materials \cite{Meiklejohn,Johnson}. Toroids made of isotropic magnetic materials
are commonplace as cores in transformers and recording heads \cite{McCaig}, and anisotropic magnetic toroids are expected
to perform better by delivering optimal spatial distributions of  $\Phi(\#r)$.

The plan of this paper is as follows. Section~\ref{bvp} presents the formulation  of the boundary-value problem and the solution procedure. Section~\ref{inrd} provides and discusses illustrative numerical results. The paper concludes with
some remarks in Sec.~\ref{conc}.
Throughout the paper, vectors are  in boldface, unit vectors are decorated by carets, dyadics are  double underlined, and column vectors as well as matrices are enclosed in square brackets.

\section{Boundary-Value Problem}\label{bvp}

\subsection{Toroidal coordinates}

\begin{figure}[H]
 \centering
\includegraphics[width=0.4\linewidth]{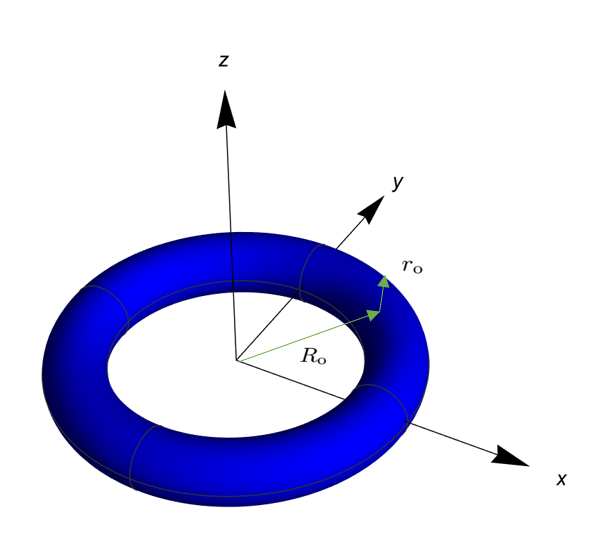}
\caption{Toroid with mean radius $\Ro$ and cross-sectional radius $\ro$.}
\label{Fig1}
\end{figure}

Figure~\ref{Fig1} shows a toroid with mean radius $\Ro$ and cross-sectional radius $\ro$, with the constraint $\Ro > \ro$. These two dimensions lead to the definition of the  two constants
\begin{equation}
\left.\begin{array}{l}
c=+\sqrt{\Ro^2-\ro^2}
\\[5pt]
a =  \text{csch}^{-1}\left( \ro/c\right)
\end{array}
\right\}
\end{equation}
that are needed to define an appropriate toroidal coordinate system $\#r\equiv(\xi,\eta,\phi)$.

Cartesian coordinates $\left\{x,y,z\right\}$ are given in terms of toroidal coordinates $\left\{\xi,\eta,\phi\right\}$ by \cite{Morse}
\begin{subequations}
\begin{equation}
x = c\frac{ \sinh \xi \cos \phi}{\cosh \xi - \cos \eta},
\end{equation}
\begin{equation}
y = c\frac{ \sinh \xi \sin \phi}{\cosh \xi - \cos \eta},
\end{equation}
and
\begin{equation}
z = c\frac{ \sin \eta}{\cosh \xi - \cos \eta}.
\end{equation}
\end{subequations}
Toroidal coordinates are given in terms of Cartesian coordinates as
\begin{subequations}
\begin{equation}
\xi = \log \left[\frac{\left( \sqrt{x^2 + y^2} + c \right)^2+z^2}{\left( \sqrt{x^2 + y^2} - c \right)^2+z^2}\right]^{1/2},
\end{equation}
\begin{equation}
\eta = \tan^{-1} \left( \frac{2 c z}{x^2 + y^2 + z^2 - c^2} \right),
\end{equation}
and
\begin{equation}
\phi = \tan^{-1} \left( \frac{y}{x} \right).
\end{equation}
\end{subequations}
The coordinate $\xi = a$ is constant on the surface of the toroid, the interior of the toroid being denoted
by
\begin{subequations}
\begin{equation}
\Vminus:\left\{\left(\xi,\eta,\phi\right)\,\vert\,\xi>a, \eta \in [0, 2\pi),   \phi \in [0, 2 \pi)\right\}
\end{equation}
and the exterior by
\begin{equation}
\Vplus:\left\{\left(\xi,\eta,\phi\right)\,\vert\,\xi<a, \eta \in [0, 2\pi),   \phi \in [0, 2 \pi)\right\}\,.
\end{equation}
\end{subequations}

\subsection{Formulation}

Let the toroid be made of an anisotropic magnetic  material  described by the constitutive relation
\begin{subequations}
\begin{equation}
\label{constitutive-one}
\#B(\#r)=\=\mu\.\#H(\#r)=
\muo\,\mur\,  \=A \.\=A \.    \#H(\#r)\,,\quad \xi > a\,,
\end{equation}
where $\muo$ denotes the permeability of free space;  the scalar $\mur>0$; and
the diagonal dyadic \cite{Chen,EAB}
\begin{equation}
\label{Aone-def}
\=A = \alpha_x^{-1}\,\ux \ux + \alpha_y^{-1} \,\uy \uy + \uz \uz
\end{equation}
\end{subequations}
contains the scalars
$\alpha_x>0$ and $\alpha_y>0$.
The exterior region $\Vplus $ is taken to be vacuous with constitutive relation
\begin{equation}
\#B(\#r)=\muo\#H(\#r)\,,\quad \xi < a\,.
\end{equation}

\subsection{Potential in  region $\Vplus$}\label{bvp-ext}
Equation~(\ref{Laplace}) applies in $\Vplus$. Its solution in the toroidal coordinate system
yields \cite{Morse,Moon}
\begin{equation}
\label{eq4}
\Phi(\#r)= \sum_{  j\in\left\{e,o\right\}}^{} \sum_{  k\in\left\{e,o\right\}}^{} \sum_{\ell =0}^{\infty} \sum_{  n=0}^{\infty} \les \calA\jkmn \Psione(\#r) + \calB\jkmn \Psitwo(\#r)
\ris \,,\quad \xi < a\,,
\end{equation}
where the toroidal basis functions
\begin{subequations}
\begin{equation}
\Psione(\#r) = \sqrt{\cosh \xi - \cos \eta}\, Q_{ n-\frac{1}{2}}^\ell (\cosh \xi) f_{ kn}(\eta)f_{  j\ell}(\phi)
\end{equation}
and
\begin{equation}
\Psitwo(\#r) = \sqrt{\cosh \xi - \cos \eta}\, P_{ n-\frac{1}{2}}^\ell (\cosh \xi)
f_{  kn}(\eta)f_{  j\ell}(\phi)
\end{equation}
\end{subequations}
involve
the trigonometric functions
\begin{equation}
\left.\begin{array}{l}
f_{ e\bar{n}}(\psi) =\cos\left(\bar{n}\psi\right)
\\[5pt]
f_{ o\bar{n}}(\psi) =\sin\left(\bar{n}\psi\right)
\end{array}
\right\}\,.
\end{equation}
The associated Legendre function of the first kind
\begin{subequations}
\begin{equation}
P_{n-\frac{1}{2}}^\ell(\cosh \xi) = \frac{i^{\ell} \Gamma\left(n+{\ell}+ \displaystyle \frac{1}{2} \right) \tanh^{\ell} \xi}{2^{\ell} {\ell}! \Gamma \left(n - {\ell} + \displaystyle \frac{1}{2} \right) \cosh^{n+\frac{1}{2}} \xi}\, _2F_1 \left( \frac{{\ell} + n + \frac{1}{2}}{2}, \frac{{\ell} + n + \frac{3}{2}}{2}, {\ell} + 1; \tanh^2 \xi  \right)
\end{equation}
and the associated Legendre function of the second kind
\begin{equation}
Q_{n-\frac{1}{2}}^\ell(\cosh \xi) = \frac{\Gamma \left( \displaystyle \frac{1}{2} \right) \Gamma\left(n+{\ell}+ \displaystyle \frac{1}{2} \right) \tanh^{\ell} \xi}{\Gamma \left( n+1 \right) 2^{n+ \frac{1}{2}} \cosh^{n+\frac{1}{2}} \xi}\, _2F_1 \left( \frac{{\ell} + n + \frac{1}{2}}{2}, \frac{{\ell} + n + \frac{3}{2}}{2}, n + 1; \text{sech}^2 \xi  \right)
\end{equation}
\end{subequations}
contain  $i=\sqrt{-1}$, $\Gamma(\.)$ as the Gamma function, and $_2F_1 (\.,\.,\.;\.)$ as the hypergeometric function \cite{Morse,Abramowitz}.
The coefficients $\calA\jkmn$ are associated
with terms that are regular as $\xi\to 0$
 whereas the coefficients $\calB\jkmn$ are associated with terms
that are regular as $\xi\to\infty$.

On the right side of Eq.~(\ref{eq4}),
\begin{equation}
\label{Phi-inc}
\Phi\inc(\#r)= \sum_{j\in\left\{e,o\right\}}^{} \sum_{k\in\left\{e,o\right\}}^{}\sum_{m=0}^{\infty} \sum_{n=0}^{\infty} \calA\jkmn \Psione(\#r)
\end{equation}
is the source potential whereas
\begin{equation}
\label{Phi-sca}
\Phi\sca(\#r)= \sum_{j\in\left\{e,o\right\}}^{} \sum_{k\in\left\{e,o\right\}}^{} \sum_{m=0}^{\infty} \sum_{n=0}^{\infty}  \calB\jkmn \Psitwo(\#r) \,,\quad \xi < a\,,
\end{equation}
is the perturbation potential.  If the toroid
were to be absent, Eq.~(\ref{eq4}) would hold for all $\#r$ with $\calB\jkmn\equiv 0$~$\forall \lec{j, k, \ell,n}\ric$.
We proceed with the assumption that the coefficients $\calA\jkmn$ are
known but  the coefficients $\calB\jkmn$ are not. Furthermore, Eq.~(\ref{Phi-inc}) is required to hold
in some sufficiently large   region that contains the toroidal region $\Vminus$ but not the region containing the source
of $\Phi\inc(\#r)$.

\subsection{Source potential}\label{sourcepot}

Sources of two types are considered for illustrative numerical results provided
in Sec.~\ref{inrd}: (i) a uniform-magnetostatic-field source and (ii) a point-magnetic-dipole source.

For the uniform-magnetostatic-field source, we have   $\vec{H}_\text{source} = -U_{sx} \ux -U_{sy} \uy -U_{sz} \uz$, with $U_{sx}$, $U_{sy}$, and $U_{sz}$ as constants, and hence the source potential is written as
\begin{equation}
\label{Uniform_Cartesian}
\Phi\inc(\#r)= U_{sx} x + U_{sy} y + U_{sz} z.
\end{equation}
On transforming Eq. (\ref{Uniform_Cartesian}) to toroidal coordinates, we obtain
\begin{equation} \label{Uniform_Toroidal}
\Phi\inc(\#r)= \frac{c}{\cosh \xi - \cos \eta} \left( U_{sx} \sinh \xi \cos \phi + U_{sy} \sinh \xi \sin \phi + U_{sz} \sin \eta \right).
\end{equation}

To find the coefficients $\calA\jkmn$ in Eq. (\ref{Phi-inc}) for the uniform-magnetostatic-field source, we equate Eqs. (\ref{Phi-inc}) and (\ref{Uniform_Toroidal}),   multiply both sides by $\Psione(\#r) \left( \displaystyle \frac{1}{\cosh \xi - \cos \eta} \right)$,   set $\xi = a$, and   integrate over $\eta \in[0,2 \pi)$ and $\phi\in[0,2\pi)$. This gives
\begin{equation}
\begin{aligned}
\calA\jkmn &=   \frac{c}{\pi^2 (1 + \delta_{{\ell}0}) (1 + \delta_{n 0}) \left[Q_{n-1/2}^{\ell}(\cosh a)\right]^2} \times \\
& \displaystyle \int_{\phi=0}^{2\pi}
\int_{\eta=0}^{2 \pi}  \left( U_{sx} \sinh \xi \cos \phi + U_{sy} \sinh \xi \sin \phi + U_{sz} \sin \eta \right)  \Psione(\#r)   \left( \displaystyle \frac{1}{\cosh \xi - \cos \eta } \right)^2
\, \Biggr|_{\xi = a} \,\mathrm{d}\eta\,\mathrm{d}\phi
 \,,
 \end{aligned}
\end{equation}
where $\delta_{{\ell}{\ell}^\prime}$ is the Kronecker delta.

For the point-magnetic-dipole source of magnetic moment $\#m = {\msx} \ux + {\msy} \uy + \msz \uz$, where ${\msx}$, $\msy$, and ${\msz}$ are constants, located at  $\bro\equiv(\xio,\etao,\phio)$, where $\xio < a$, $\etao \in[0,2 \pi)$, and $\phio \in[0,2\pi)$, the source coefficients can be obtained with the use of the bilinear expansion of the Green function in toroidal coordinates as \cite{Bates}
\begin{equation}
\label{Atilde-ps-d}
\calA\jkmn  = \frac{ \muo {|\vec{m}|}}{4 \pi^2 c} \frac{\left(2-\delta_{{\ell}0}\right)\left(2-\delta_{n0} \right)(-1)^m \Gamma \left(n-{\ell}+\frac{1}{2} \right)}{\Gamma \left(n+{\ell}+\frac{1}{2} \right)} \, \um \. \nabla_{\rm o}\left[\Psitwo(\bro) \right],
\end{equation}
where ${|\vec{m}|}>0$ and the unit vector $\um =\#m/|\vec{m}|$ defines the orientation of the point magnetic dipole.
See the Appendix for additional details.

\subsection{Potential in  region $\Vminus$}\label{bvp-core}
Inside the magnetic toroid, the potential $\Phi(\#r)$ does not obey the Laplace equation;
instead,
\begin{equation}
\label{eq8-1}
\nabla\.\les   \= A \.\=A \.  \nabla\Phi(\#r)\ris=0\,,\quad \xi>a\,.
\end{equation}
In order to solve this equation, let us make an affine  coordinate transformation from $\#r$
to $\#r_{1}\equiv(\xi_{1},\eta_{1},\phi_{1})$ as follows \cite{QJMAM1}:
\begin{equation}
\label{transf-1}
\#r_{1}=  \=A^{-1}\.  \#r\,.
 \end{equation}
Since $\alpha_x>0$ and $\alpha_y>0$,
the bijective transformation (\ref{transf-1}) maps a sphere into an ellipsoid.
That ellipsoid's principal axes are the same as
 the  Cartesian unit vectors affixed to the sphere.
 In such a situation, the variables $(\xi_{1},\eta_{1},\phi_{1})$ are given by
\begin{equation}
\begin{rcases}
\begin{aligned}
\xi_1(\#r) &=  \log  \left\{ \frac{\displaystyle \frac{ \sin^2 \eta}{(\cosh \xi - \cos \eta)^2} + \left[ 1 + \frac{ \sinh \xi\sqrt{\alpha_x^2 \cos^2 \phi + \alpha_y^2 \sin^2 \phi} }{\cosh \xi - \cos \eta} \right]}{\displaystyle \frac{ \sin^2 \eta}{(\cosh \xi - \cos \eta)^2} + \left[ 1 - \frac{ \sinh \xi\sqrt{\alpha_x^2 \cos^2 \phi + \alpha_y^2 \sin^2 \phi} }{\cosh \xi - \cos \eta} \right]} \right\}^{1/2} \\
\eta_1(\#r) &= \tan^{-1} \left[ \frac{2 \sin \eta \, (\cosh \xi - \cos \eta)}{\sin^2 \eta + \sinh^2 \xi \, (\alpha_x^2 \cos^2 \phi + \alpha_y^2 \sin^2 \phi) - (\cosh \xi - \cos \eta)^2} \right]  \\
\phi_1(\#r) & = \tan^{-1} \left( \frac{\alpha_y}{\alpha_x} \tan \phi \right)
\end{aligned}
\end{rcases}.
\end{equation}

In consequence of Eq.(\ref{transf-1}), Eq.~(\ref{eq8-1}) is transformed to
\begin{equation}
\label{eq9-1}
  \nabla_{1}^2\,\Phi(\#r_{1}) =0\,,
\end{equation}
which is the Laplace equation in the transformed space.
Its solution is given by \cite{Morse,Moon}
\begin{equation}
\label{eq10-1}
\Phi(\#r_{1})= \sum_{j\in\left\{e,o\right\}}^{} \sum_{k\in\left\{e,o\right\}}^{} \sum_{m=0}^{\infty} \sum_{n=0}^{\infty} \les \calC\jkmn \Psione(\#r_1) + \calD\jkmn \Psitwo(\#r_1)
\ris\,.
\end{equation}
We must set $\calD\jkmn\equiv0$ in order
to exclude terms on the right side
of Eq.~(\ref{eq10-1}) that are not regular as $\xi\to\infty$.

Accordingly, we define the potential inside the toroid as
\begin{equation}
\label{eq11-1}
\Phi_\text{int}(\#r)=  \sum_{j\in\left\{e,o\right\}}^{} \sum_{k\in\left\{e,o\right\}}^{} \sum_{m=0}^{\infty} \sum_{n=0}^{\infty} \calC\jkmn \Psione(\#r_1) \,,\quad \xi > a\,.
\end{equation}
The unknown coefficients $\calC\jkmn$ have to be determined.

\subsection{Boundary conditions}\label{bvp-bc}
With the assumption of the interface $\xi=a$ being current-free, the tangential component of the magnetostatic field must be continuous across the interface $\xi=a$ \cite{Chen}. Therefore, the
potential must be continuous across both interfaces; hence,
\begin{equation}
\label{bc1-2}
\Phi\inc(\xi,\eta,\phi) + \Phi\sca(\xi,\eta,\phi)=\Phi_\text{int}(\xi,\eta,\phi)\,,
\quad \xi=a\,,\quad
\eta\in[0,2\pi)\,,\quad \phi\in[0,2\pi)\,.
\end{equation}

As the normal component of the
magnetostatic flux density must be continuous across the interface $\xi=a$ \cite{Chen}, we get
\begin{eqnarray}
\nonumber
&&
\uxi \. \nabla \les\Phi\inc(\xi,\eta,\phi) + \Phi\sca(\xi,\eta,\phi)\ris=
\mur\,\uxi\.  \=A\.\=A\. \nabla
\Phi_\text{int}(\xi,\eta,\phi)\,,
\\[5pt]
&&
\qquad\qquad \xi=a\,,\quad
\eta\in[0,2 \pi)\,,\quad \phi\in[0,2\pi)\,,
\label{bc2-2}
\end{eqnarray}
where the unit vector $\uxi=\displaystyle \frac{1- \cosh \xi \cos \eta}{\cosh \xi - \cos \eta} (\ux\cos\phi+\uy\sin\phi) -\uz \frac{\sinh \xi \sin \eta}{\cosh \xi - \cos \eta}$.

\subsection{Transition Matrix}\label{transmat}

On multiplying both sides of Eqs.~(\ref{bc1-2}) and (\ref{bc2-2}) by $\Psionep(\#r) \left( \displaystyle \frac{1}{\cosh \xi - \cos \eta} \right)$, setting $\xi = a$, and integrating over $\eta \in[0,2 \pi)$
and $\phi\in[0,2\pi)$, the following sets of simultaneous algebraic equations are obtained:
\begin{equation} \label{eq1}
\sum_{jk{\ell}n} \left( \calI\double \calA\jkmn + \calJ\double \calB\jkmn  \right) = \sum_{jk{\ell}n} \calK\double \calC\jkmn,
\end{equation}
and
\begin{equation} \label{eq2}
\sum_{jk{\ell}n} \left( \calL\double \calA\jkmn + \calM\double \calB\jkmn  \right) =  \sum_{jk{\ell}n} \calN\double \calC\jkmn,
\end{equation}
where
\begin{equation}
\calI\double = \pi^2 (1 + \delta_{{\ell}0}) (1 + \delta_{n 0}) \left[Q_{n-1/2}^{\ell}(\cosh a)\right]^2 \delta_{j j'} \delta_{k k'} \delta_{{\ell}{\ell}'} \delta_{n n'},
 \label{eq-I}
\end{equation}
\begin{equation}
\calJ\double = \pi^2 (1 + \delta_{{\ell}0}) (1 + \delta_{n 0}) P_{n-1/2}^{\ell}(\cosh a) Q_{n-1/2}^{\ell}(\cosh a) \delta_{j j'} \delta_{k k'} \delta_{{\ell}{\ell}'} \delta_{n n'},
 \label{eq-J}
\end{equation}
\begin{equation}
\calK\double = \int_{\phi=0}^{2\pi}
\int_{\eta=0}^{2 \pi}  \Psione(\#r_1)  \Psionep(\#r)    \left( \displaystyle \frac{1}{\cosh \xi - \cos \eta} \right)
\,  \Biggr|_{\xi = a} \,\mathrm{d}\eta\,\mathrm{d}\phi
 \,,
  \label{eq-K}
\end{equation}
\begin{equation}
\calL\double =  \int_{\phi=0}^{2\pi}
\int_{\eta=0}^{2 \pi}  \uxi \. \nabla \left[\Psione(\#r)\right]  \Psionep(\#r)    \left( \displaystyle \frac{1}{\cosh \xi - \cos \eta} \right)
\,  \Biggr|_{\xi = a} \,\mathrm{d}\eta\,\mathrm{d}\phi
 \,,
  \label{eq-L}
\end{equation}
\begin{equation}
\calM\double =  \int_{\phi=0}^{2\pi}
\int_{\eta=0}^{2 \pi}  \uxi  \. \nabla \left[\Psitwo(\#r)\right]  \Psionep(\#r)    \left( \displaystyle \frac{1}{\cosh \xi - \cos \eta} \right)
\,  \Biggr|_{\xi = a} \,\mathrm{d}\eta\,\mathrm{d}\phi
 \,,
  \label{eq-M}
\end{equation}
and
\begin{equation}
\calN\double = \mur \int_{\phi=0}^{2\pi}
\int_{\eta=0}^{2 \pi}  \uxi \.
\=A \. \=A \.
 \nabla \left[\Psione(\#r_1)\right]  \Psionep(\#r)    \left( \displaystyle \frac{1}{\cosh \xi - \cos \eta} \right)
\,  \Biggr|_{\xi = a} \,\mathrm{d}\eta\,\mathrm{d}\phi
 \,.
 \label{eq-N}
\end{equation}

After  truncating the indexes ${\ell}$, ${\ell}^\prime$, $n$, and $n^\prime$ so that only ${\ell}\in[0,N]$, ${\ell}^\prime\in[0,N]$, $n\in[0,N]$, and $n^\prime\in[0,N]$ are considered with $N>0$,
Eqs.~(\ref{eq1}) and (\ref{eq2}) can be put together  in matrix form symbolically as
\begin{equation}
\label{m_eq1}
\les\calI\ris\,\les\calA\ris +\les\calJ\ris\,\les\calB\ris=\les\calK\ris\,\les\calC\ris
\end{equation}
and
\begin{equation}
\label{m_eq2}
\les\calL\ris\,\les\calA\ris +\les\calM\ris\,\les\calB\ris= \les\calN\ris\,\les\calC\ris\,.
\end{equation}

The system of Eqs.~(\ref{m_eq1}) and (\ref{m_eq2}) can be written as
\begin{equation} \label{ABC-def}
\begin{bmatrix}
\les\calA\ris  \\ \les\calB\ris
\end{bmatrix}
= \begin{bmatrix}
\les\calS\ris  \\ \les\calU\ris
\end{bmatrix}
 \les\calC\ris,
\end{equation}
where
\begin{equation}
\label{SU-def}
\begin{bmatrix}
\les\calS\ris  \\ \les\calU\ris
\end{bmatrix}
 =
 \begin{bmatrix}
\les\calI\ris && \les\calJ\ris  \\ \les\calL\ris && \les\calM\ris
\end{bmatrix}^{-1}
\begin{bmatrix}
\les\calK\ris  \\ \les\calN\ris
\end{bmatrix}.
\end{equation}
Then, the   column vector $\les\calB\ris$ containing the coefficients in the truncated
series representing the perturbation potential is related to the column vector $\les\calA\ris$
containing the coefficients in the truncated
series representing the source potential  by
\begin{equation}
\label{T-def}
 \les\calB\ris =\les\calT\ris\,\les\calA\ris\,.
 \end{equation}
 In this equation,
  the transition matrix $\les\calT\ris$, capturing the perturbational characteristics of the anisotropic  magnetic toroid, is given by
\begin{equation}
\label{Tmat}
\les\calT\ris= \les\calU\ris \les\calS\ris^{-1}
\end{equation}

\section{Illustrative Numerical Results and Discussion}\label{inrd}

\subsection{Preliminaries}

A Mathematica\texttrademark~program was written to calculate the transition matrix $\les\calT\ris$ of the anisotropic magnetic toroid
for a chosen truncation order $N$. The column vector $\les\calB\ris$ containing
the perturbation-potential coefficients was then calculated using Eq.~(\ref{T-def}). Also, the column vector $\les\calC\ris$ containing the
internal-potential coefficients was calculated using Eq.~(\ref{ABC-def}). Thereafter, the perturbation potential and the internal potential were calculated using Eqs. (\ref{Phi-sca}) and (\ref{eq11-1}), respectively. Finally, the magnetic field everywhere was calculated using
\begin{equation}
\vec{H}(\xi,\eta,\phi) = \left\{ \begin{array}{ll}
- \nabla \left[ \Phi\inc(\xi,\eta,\phi) + \Phi\sca(\xi,\eta,\phi) \right], & \xi < a,       \\[5pt]
- \nabla  \Phi_\text{int}(\xi,\eta,\phi) , & \xi > a .
\end{array} \right.
\end{equation}

Validation of the program was performed by solving independently the problem of a toroid composed of an isotropic magnetic material and implementing this solution in a separate Mathematica\texttrademark~program. Comparison between the original program (for the anisotropic magnetic toroid) against the program for the isotropic magnetic toroid was done by setting in the original program $\alpha_x = \alpha_y = 1$, with $a$ being arbitrary. Excellent agreement in the results of the two programs was obtained for several different values of $\mur$, $\Ro$, and $\ro$. Furthermore, on the toroidal surface, the source-potential expansions for the cases of the uniform-magnetostatic-field source   and the point-magnetic-dipole source  were compared against their closed-form expressions and  excellent agreements were obtained.

In general, the perturbation and internal potentials depend on the
\begin{itemize}
\item mean radius $\Ro$ and cross-sectional radius $\ro$;
\item permeability $\mur$;
\item anisotropy parameters $\alpha_x$ and $\alpha_y$; and
\item source potential characteristics.
\end{itemize}
We chose
\begin{equation}
\mur = \frac{3 \, \mu_\text{ave}}{\alpha_x^{-2}+ \alpha_y^{-2} + 1},
\end{equation}
where $\mu_\text{ave} = 500$.
Note that $\mur=\mu_\text{ave}$ when the toroid is made of an isotropic material (i.e., $\=A=\=I$, and hence $\alpha_x=\alpha_y=1$). Also, we set $\Ro/\ro =  {5/3}$ and $\ro =  {3}$ cm. For later use, we define $\tilde{R} = \Ro + \ro$.

A convergence test was carried out with respect to $N$, by calculating the perturbation and internal potentials at several points in space as $N$ was incremented by unity. The iterative process of increasing $N$ was terminated when the perturbation and internal potentials converged within a preset tolerance of $1\%$. To demonstrate the convergence procedure, we defined the  relative error
\begin{equation}
\Upsilon(N, \#r) =  \left\vert \frac{\Phi\sca(N, \#r)- \Phi\sca(N-1, \#r)}{\Phi\sca(N, \#r)}\right\vert \,.
\end{equation}
We show  $\Upsilon(N, \#r)$ as well as the computational time  in relation to $N\in[0,6]$   in Table \ref{Table1},
for a toroid made of a biaxial magnetic medium with $\alpha_x = 1.1$ and $\alpha_y = 1.2$. These calculations were performed
on the  United Arab Emirates University High-Performance Computing system consisting of three nodes, each of which is equipped with Intel Xeon E5-2690 v4 $@$ 2.60 GHz and 256~GB RAM. The source was chosen to be a point magnetic dipole located at the origin (i.e., $\xio = 0$ and $\etao = \pi$) and oriented along the $z$ axis (i.e., $\hat{\#m} = \uz$),  while the observation point was selected to be  ${\#r}\equiv\{\xi = 0.9 a, \eta = 1, \phi = 1 \}$. Per Table \ref{Table1},   $N=6$ is  sufficient for satisfying the aforementioned convergence criterion. The corresponding procedure was also applied to the internal potential, and the results were found to be quantitatively similar.
 \begin{table}
\caption{\label{Table1}
			{\bf   Relative error $\Upsilon(N, \#r)$ and computational time  to determine $\Phi\sca(N, \#r)$ as a function of $N$ 
			for ${\#r}\equiv\{\xi = 0.9 a, \eta = 1, \phi = 1 \}$,
			when a toroid made of an anisotropic material ($\mu_\text{ave} = 500$, $\alpha_x = 1.1$, $\alpha_y = 1.2$)
			 is exposed to a point magnetic dipole ${\#m} \parallel \uz$ 	located at the origin.}}
 \begin{center}
\begin{tabular}{|c| c| c|}
 \hline
  $N$ &  $\Upsilon(N, \#r)$   &Computational time  \\
  &    ($\%$) &  (min) \\[0.5ex]
 \hline\hline
 0 & - & 0.133   \\
 \hline
 1 & 100 & 6    \\
 \hline
 2 & 93 & 48   \\
 \hline
 3 & 4.86 & 150   \\
 \hline
  4 &  8.86 & 514   \\
 \hline
  5 & 3.88 & 1080   \\
 \hline
 6 &  0.31 &2055   \\  [1ex]
 \hline
\end{tabular}
\end{center}
 \end{table}

Next, we present numerical results for the variations of the magnetic field everywhere for different anisotropy parameters $\alpha_x$ and $\alpha_y$ of the toroid. The excitation is due to either a uniform-magnetostatic-field source or a point-magnetic-dipole source.

\subsection{Uniform-magnetostatic-field source}

\begin{figure}[H]
 \centering
\includegraphics[width=0.4\linewidth]{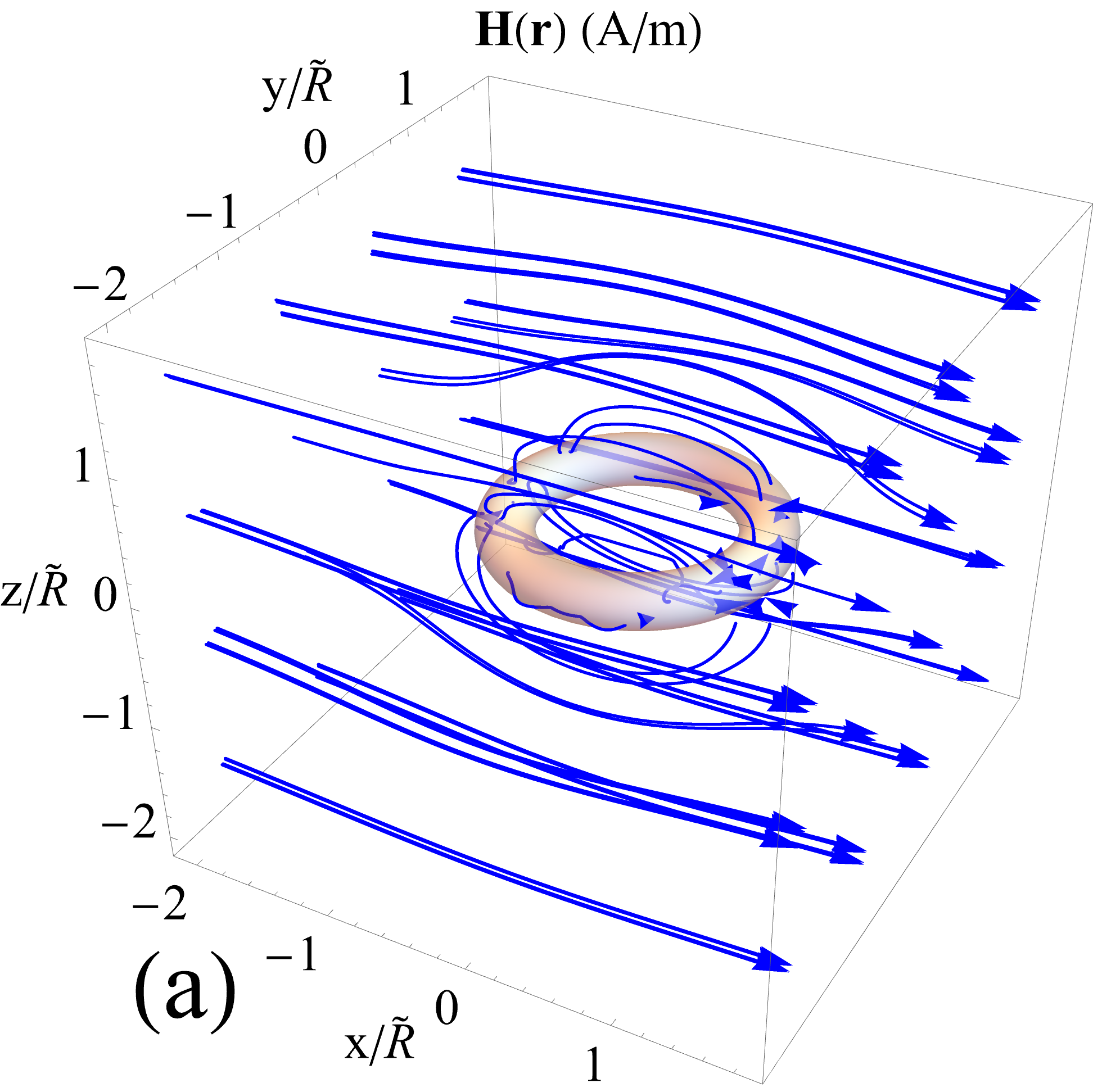}
\includegraphics[width=0.4\linewidth]{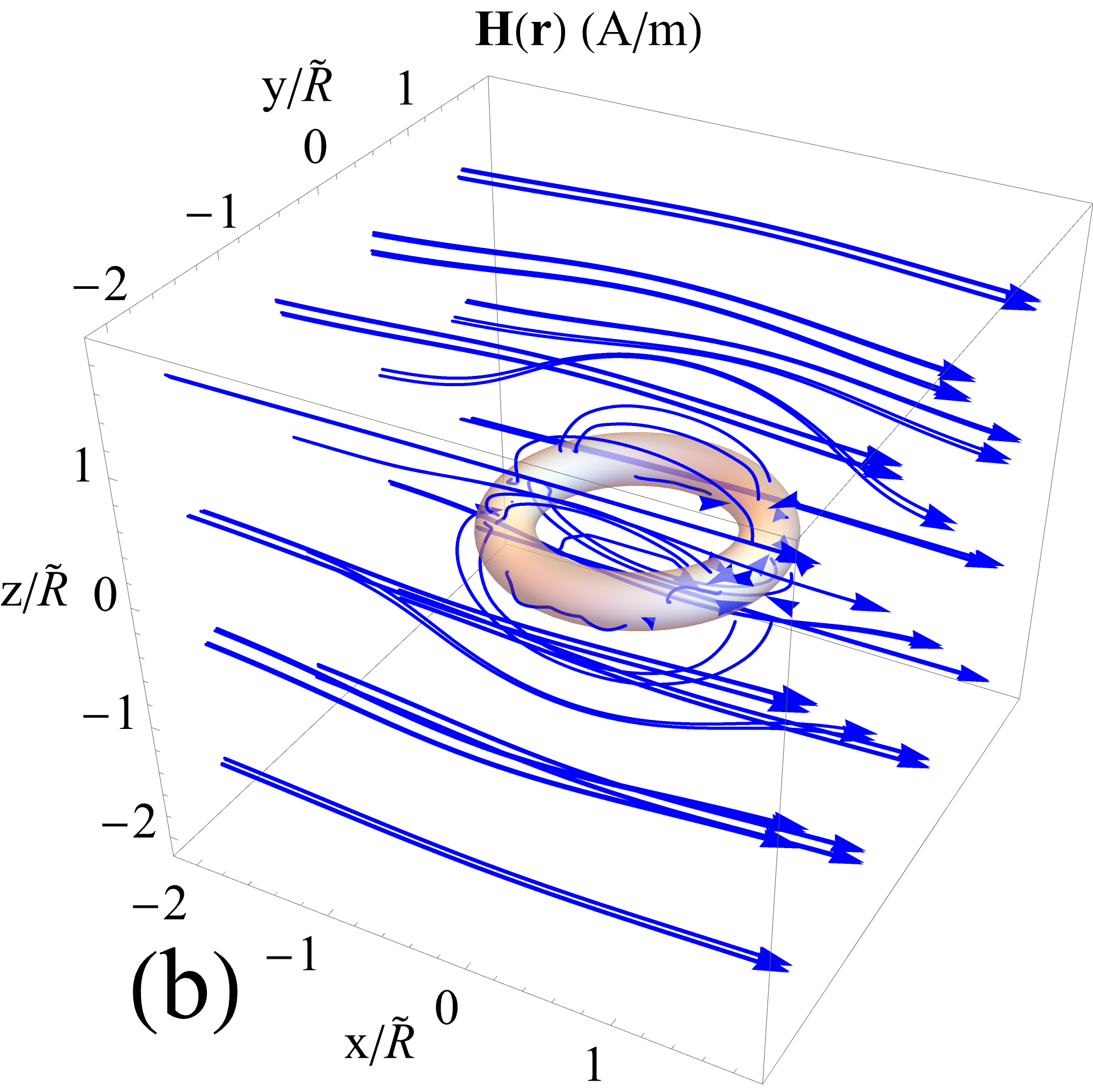}
\includegraphics[width=0.4\linewidth]{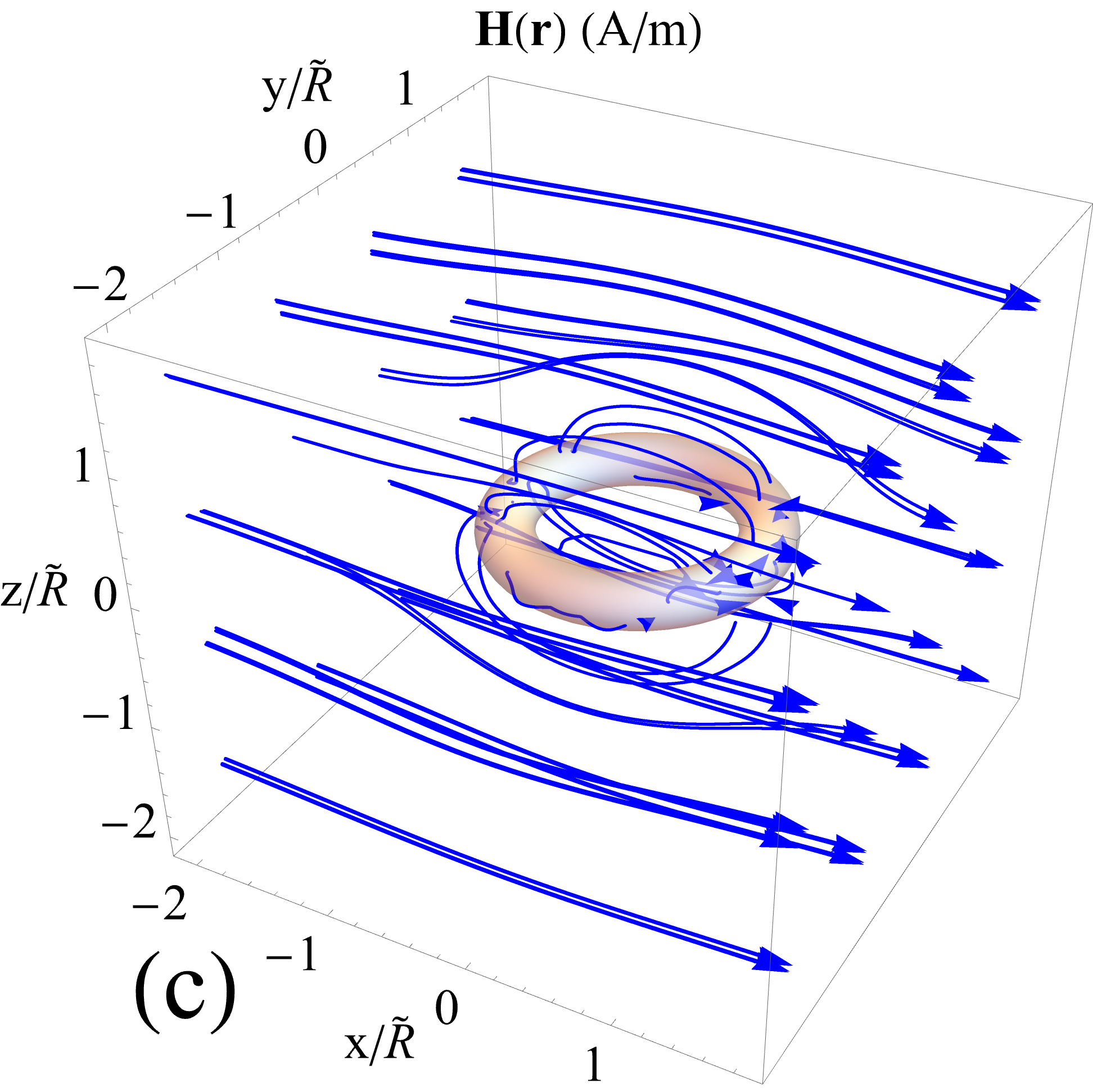}
\caption{$\vec{H}(x, y, z)$ versus $x \in [-2, 2] \tilde{R}$, $y \in [-2, 2] \tilde{R}$, and $z \in [-2, 2] \tilde{R}$ for $\vec{H}_\text{source} =  \ux$ A~m$^{-1}$ when  (a)  $\alpha_x = 1$ and $\alpha_y = 1$, (b) $\alpha_x = 1.2$ and $\alpha_y = 1.2$, and (c) $\alpha_x = 1.1$ and $\alpha_y = 1.2$.}
\label{Fig2}
\end{figure}

\begin{figure}[H]
 \centering
\includegraphics[width=0.4\linewidth]{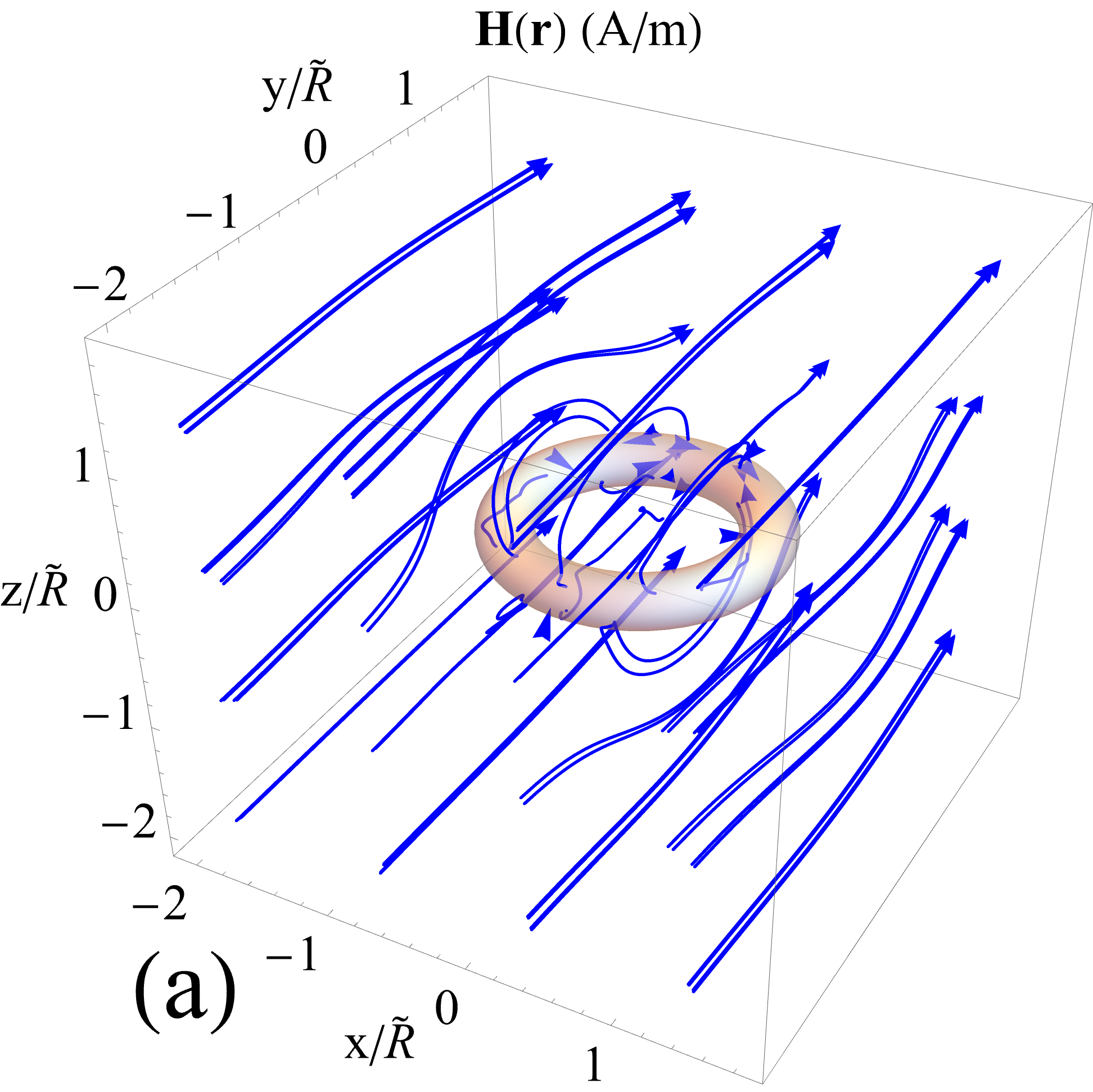}
\includegraphics[width=0.4\linewidth]{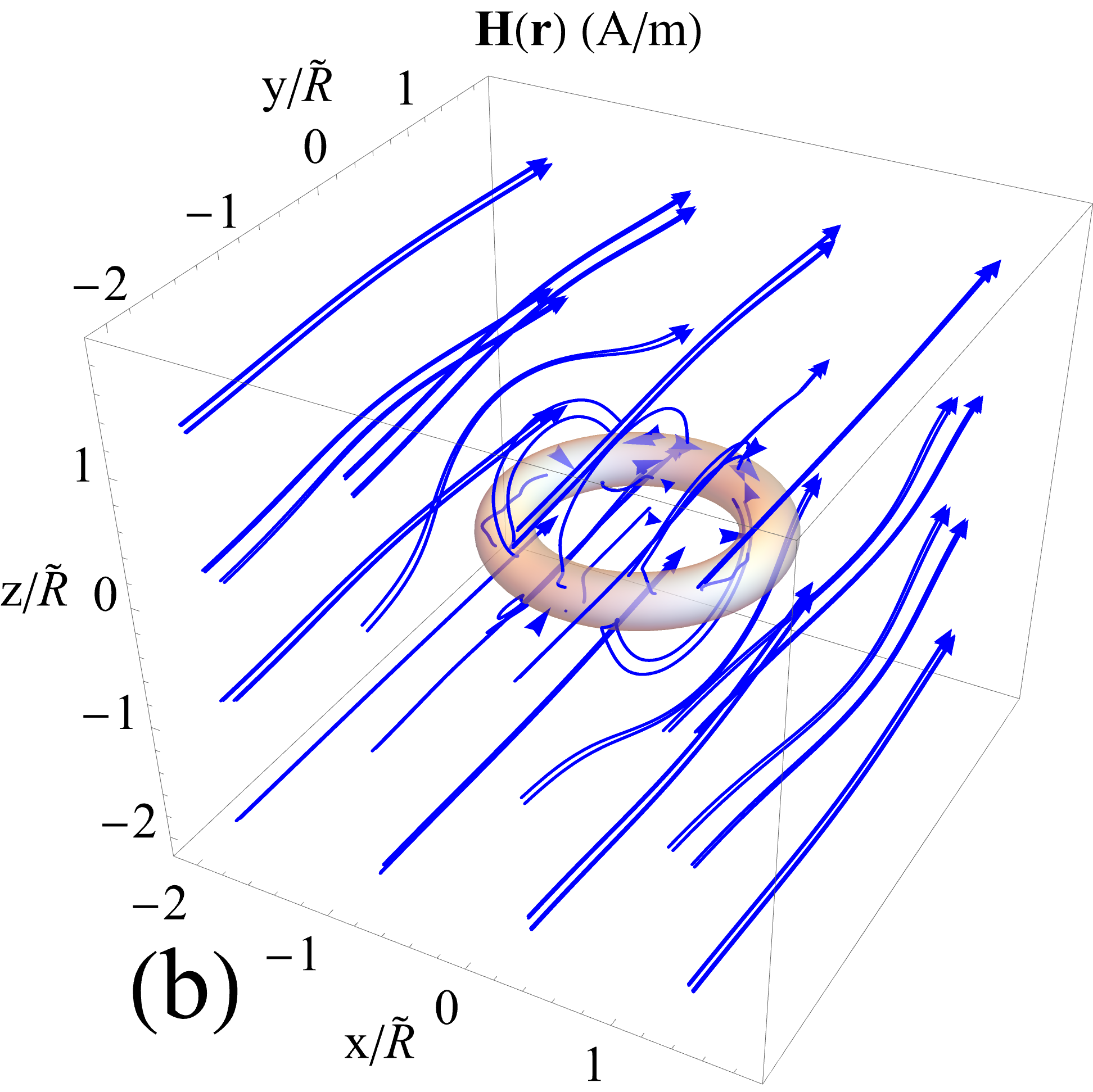}
\includegraphics[width=0.4\linewidth]{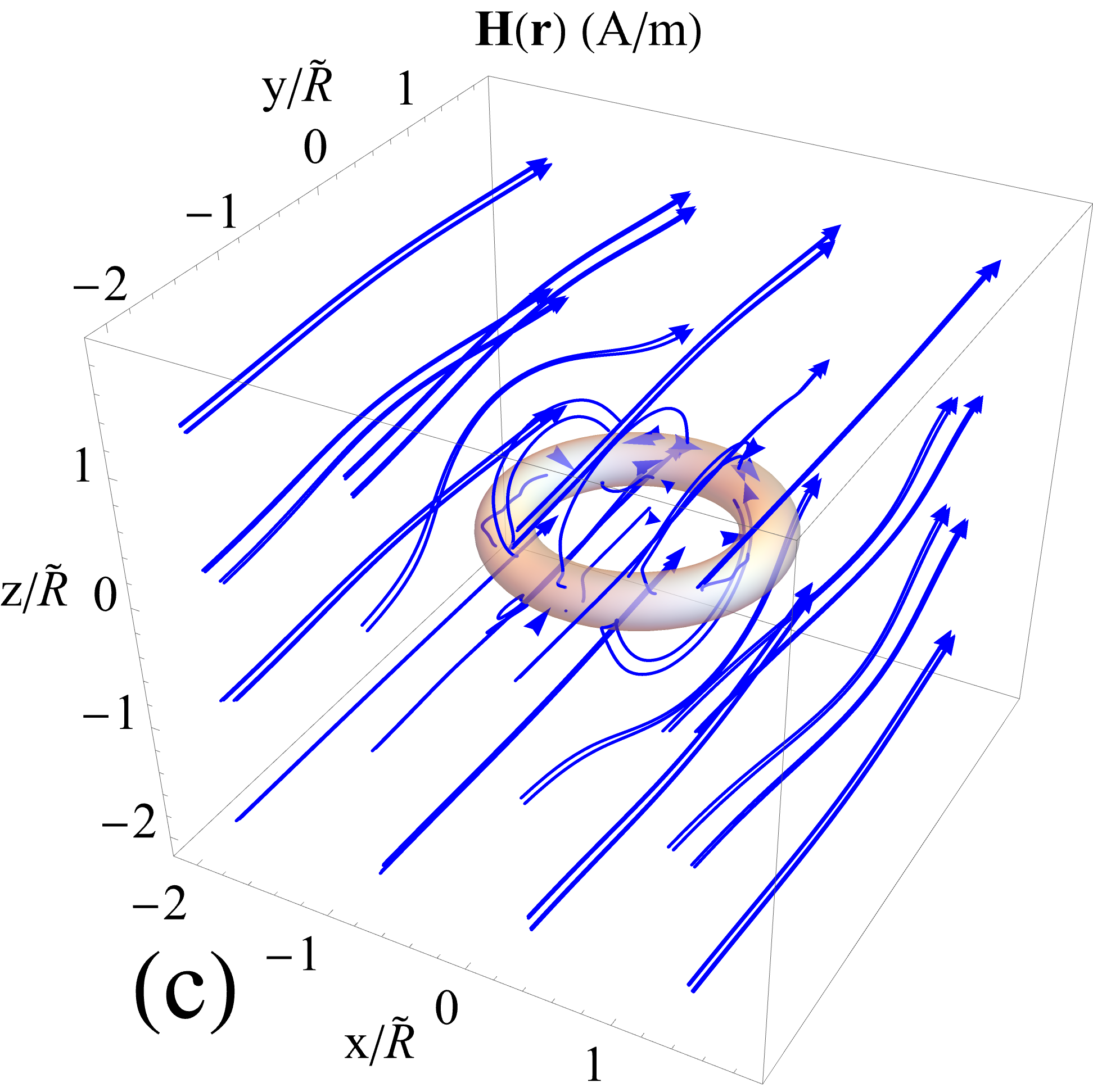}
\caption{As in Fig.~\ref{Fig2}, but for $\vec{H}_\text{source} =  \uy$ A~m$^{-1}$.}
\label{Fig3}
\end{figure}

\begin{figure}[H]
 \centering
\includegraphics[width=0.4\linewidth]{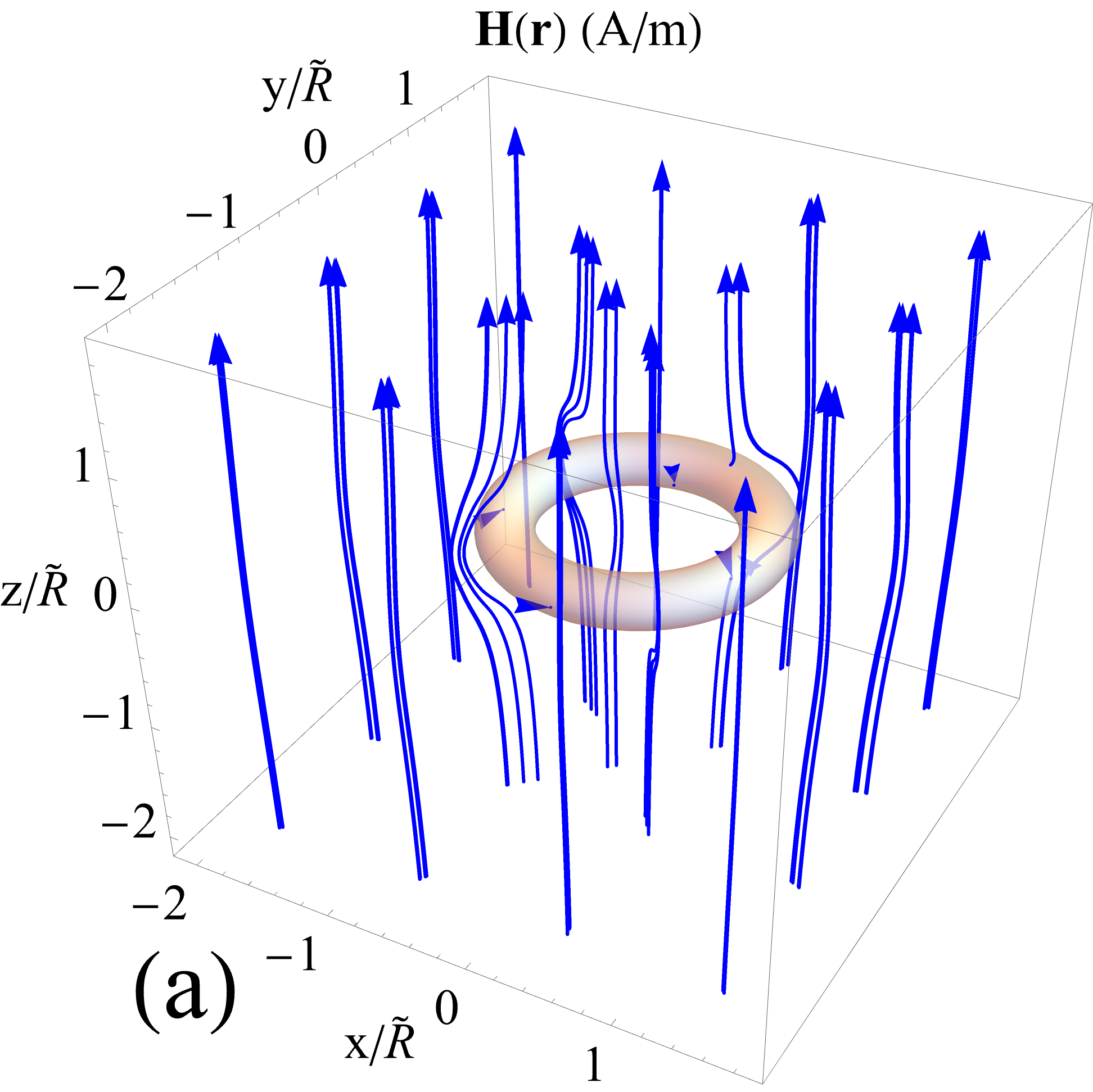}
\includegraphics[width=0.4\linewidth]{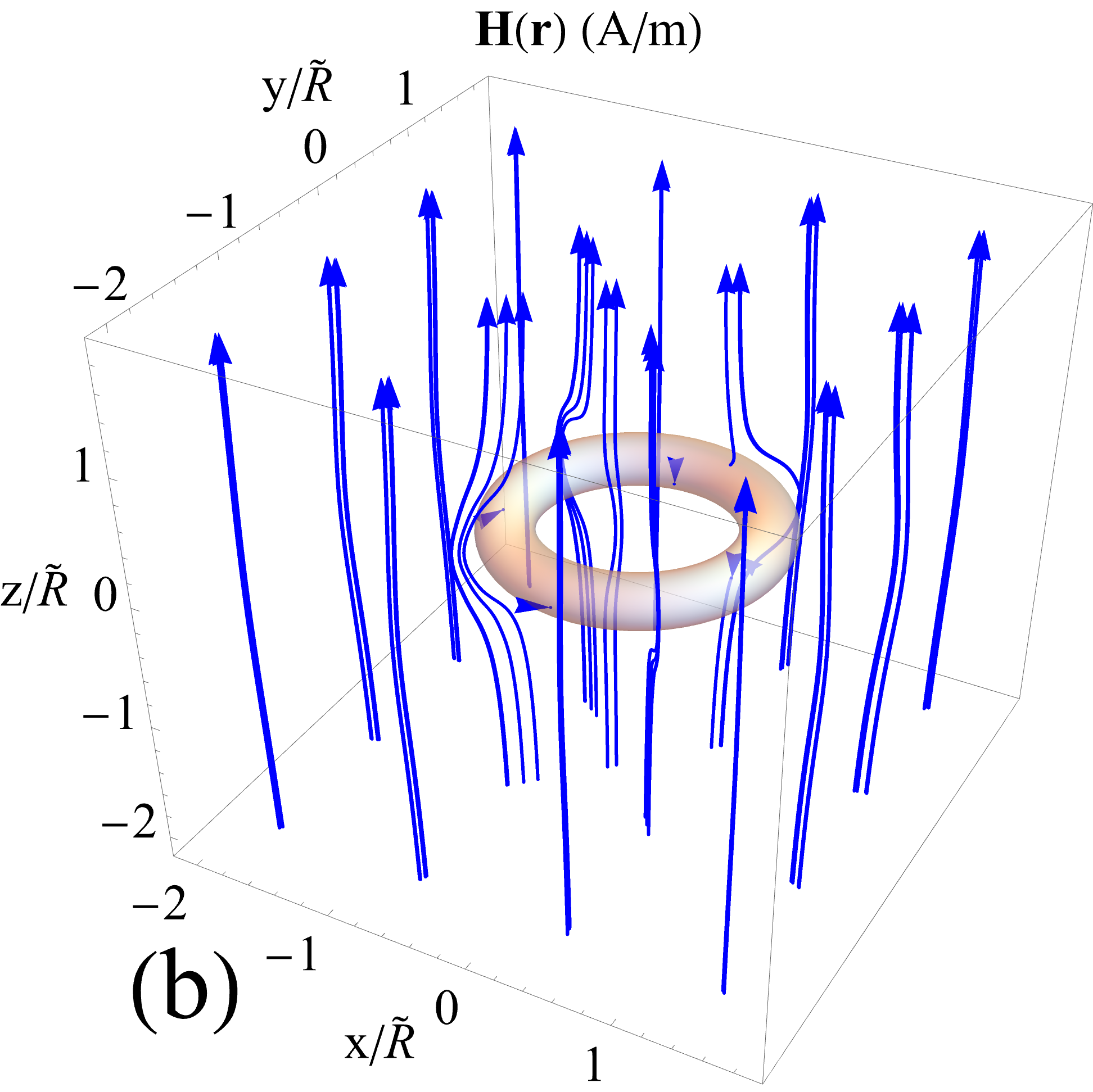}
\includegraphics[width=0.4\linewidth]{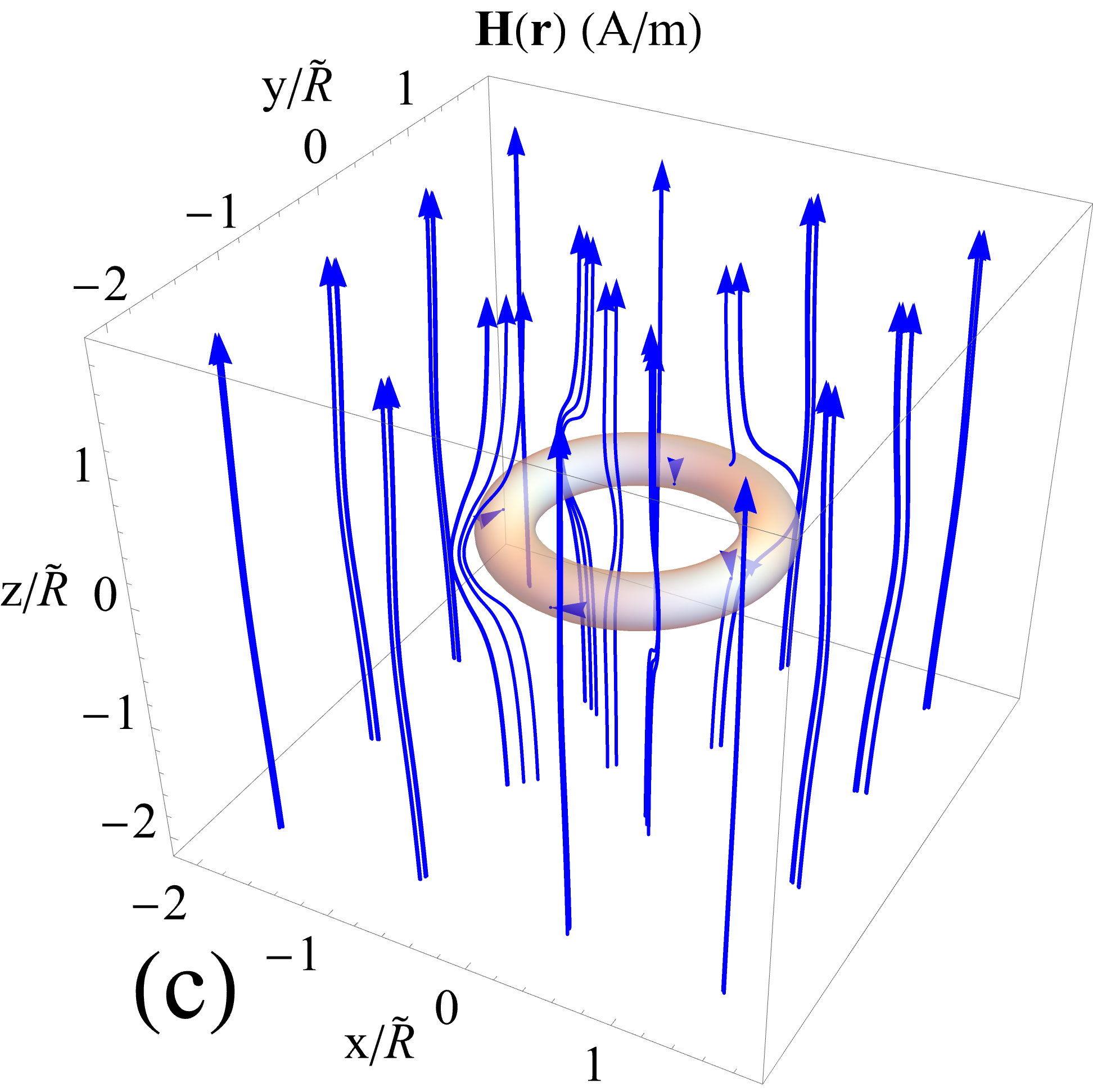}
\caption{As in Fig.~\ref{Fig2}, but for $\vec{H}_\text{source} =  \uz$ A~m$^{-1}$.}
\label{Fig4}
\end{figure}

Figures~\ref{Fig2}, \ref{Fig3}, and  \ref{Fig4} correspond to the cases $\vec{H}_\text{source} =  \ux$ A~m$^{-1}$, $\vec{H}_\text{source} =  \uy$ A~m$^{-1}$ and $\vec{H}_\text{source} =  \uz$ A~m$^{-1}$, respectively.
In each of these three figures, vector plots of the total magnetostatic field $\vec{H}$
versus $x \in [-2, 2] \tilde{R}$, $y \in [-2, 2] \tilde{R}$, and $z \in [-2, 2] \tilde{R}$  are presented, with the different panels corresponding to (a) $\alpha_x = 1$ and $\alpha_y = 1$, (b) $\alpha_x = 1.2$ and $\alpha_y = 1.2$, and (c) $\alpha_x = 1.1$ and $\alpha_y = 1.2$. Near the toroid, the direction of the total magnetostatic field changes significantly when $\vec{H}_\text{source}$ is parallel to the $x$ and $y$ axes, but exhibits only minor variations when $\vec{H}_\text{source}$ is parallel to the $z$ axis. By close inspection, one can also notice that the direction of the total magnetostatic field on the toroidal surface is affected by the change in the anisotropy parameters $\alpha_x$ and $\alpha_y$.

\subsection{Point-magnetic-dipole source}

\begin{figure}[H]
 \centering
\includegraphics[width=0.4\linewidth]{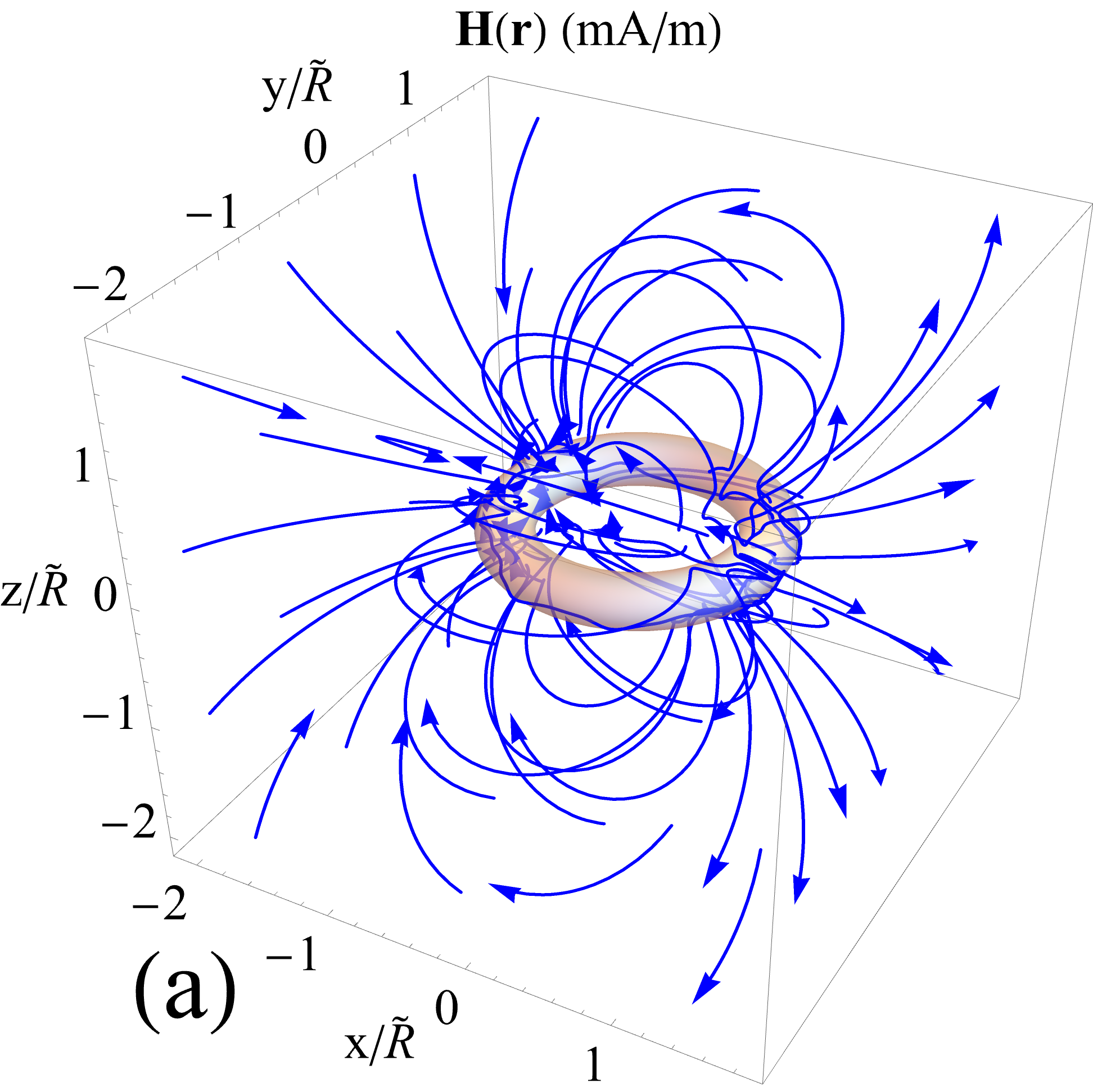}
\includegraphics[width=0.4\linewidth]{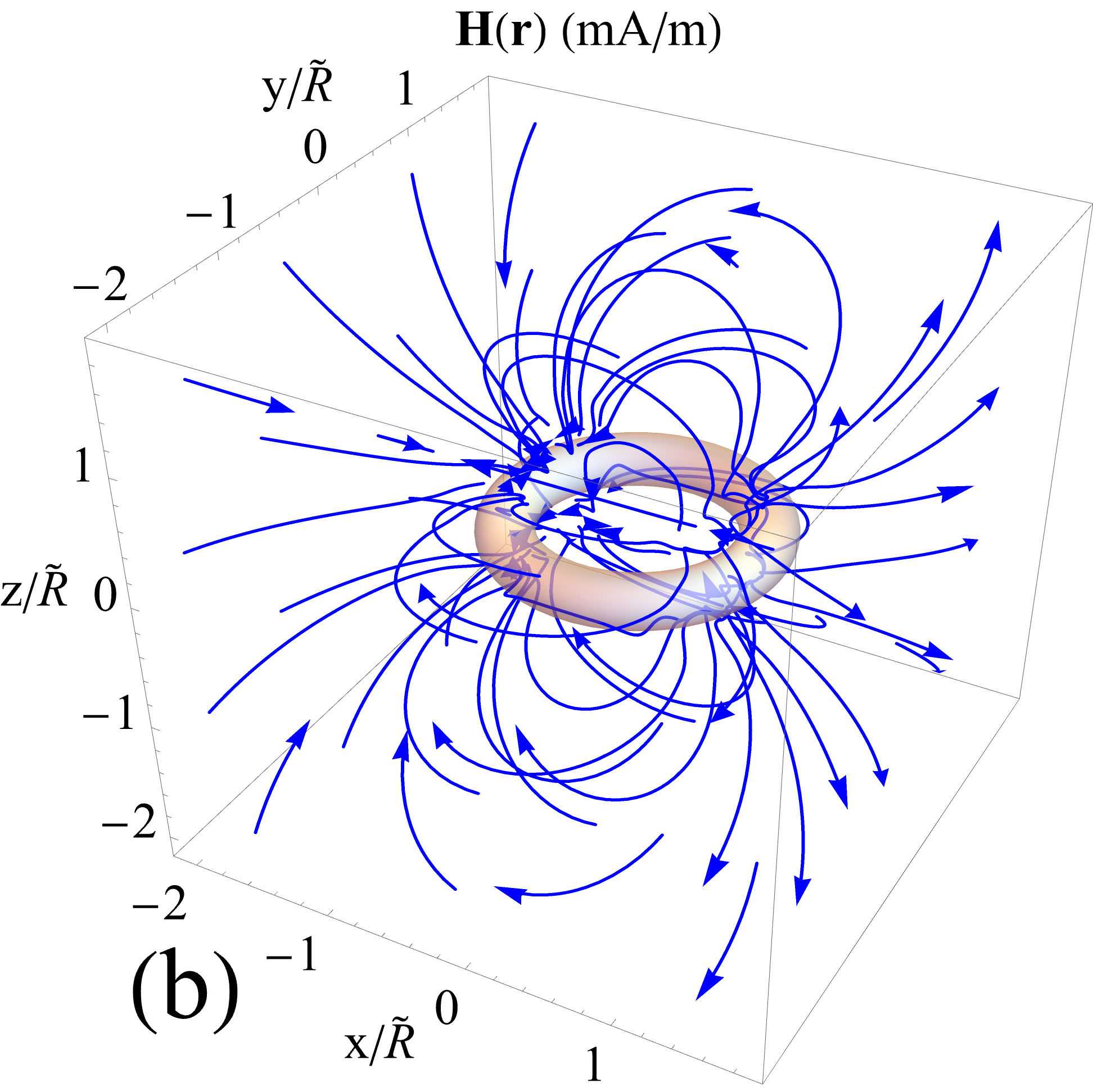}
\includegraphics[width=0.4\linewidth]{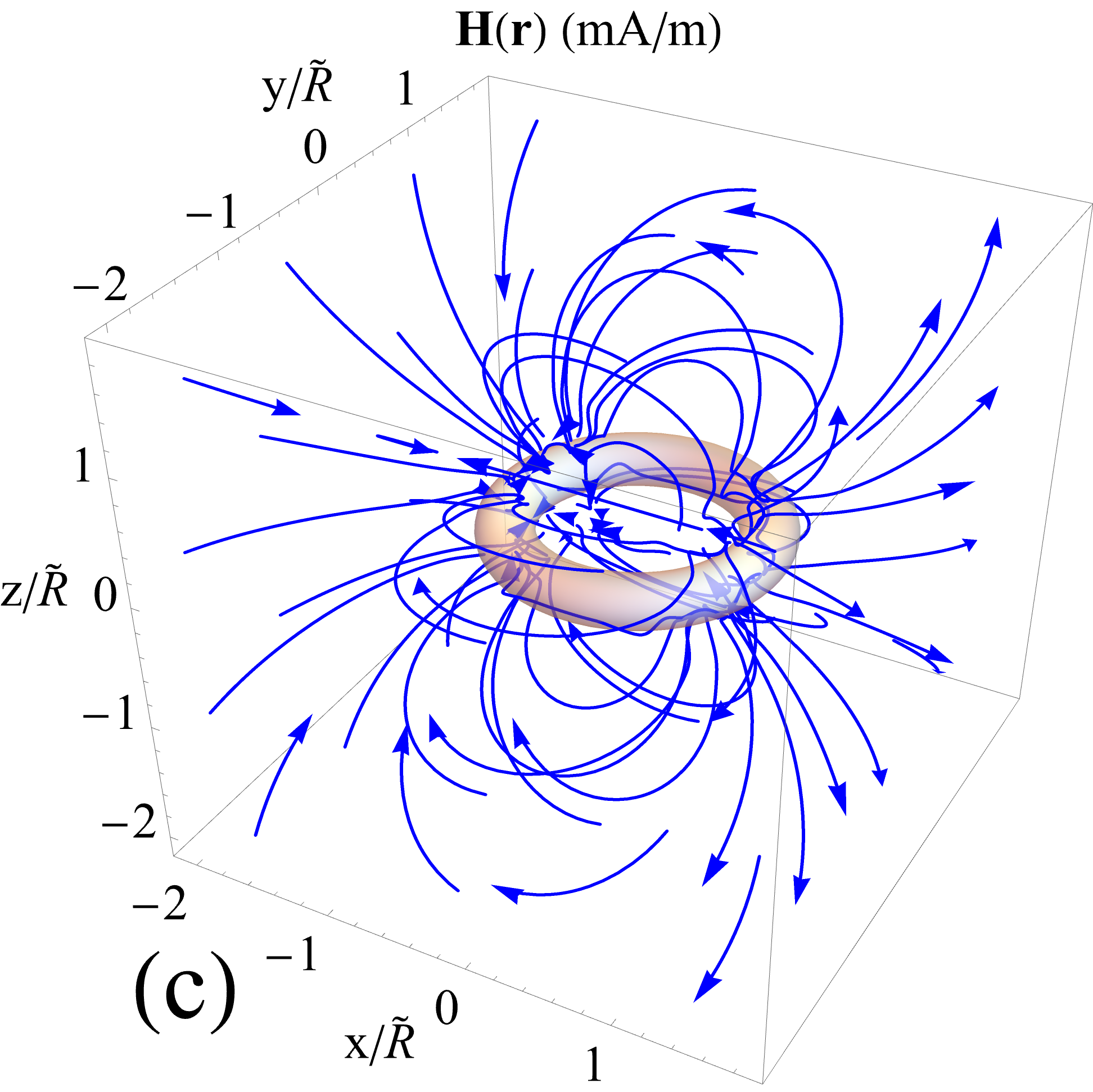}
\caption{As in Fig. \ref{Fig2}, but for a point-magnetic-dipole source with $\#m = \ux$.}
\label{Fig5}
\end{figure}

\begin{figure}[H]
 \centering
\includegraphics[width=0.4\linewidth]{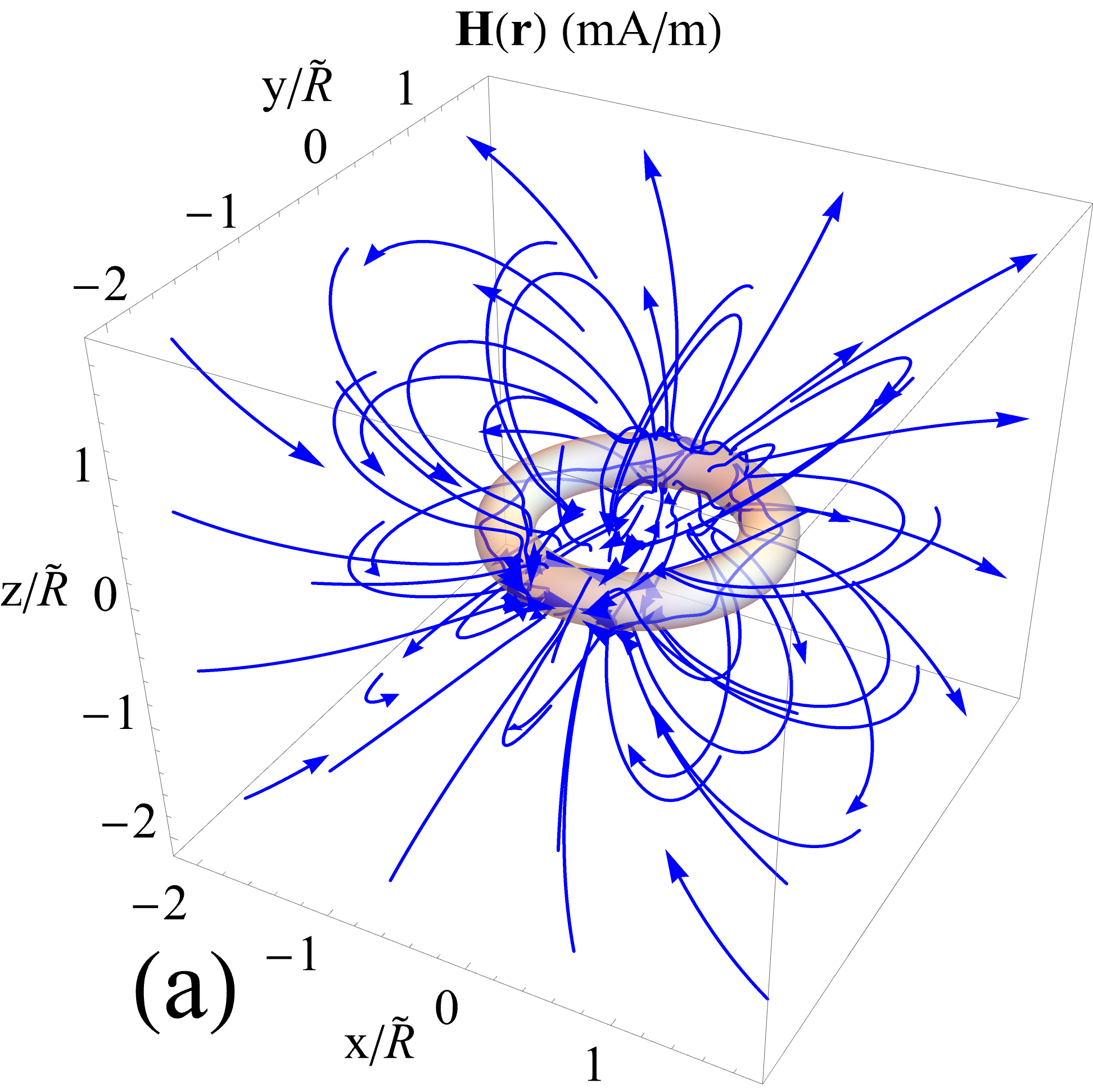}
\includegraphics[width=0.4\linewidth]{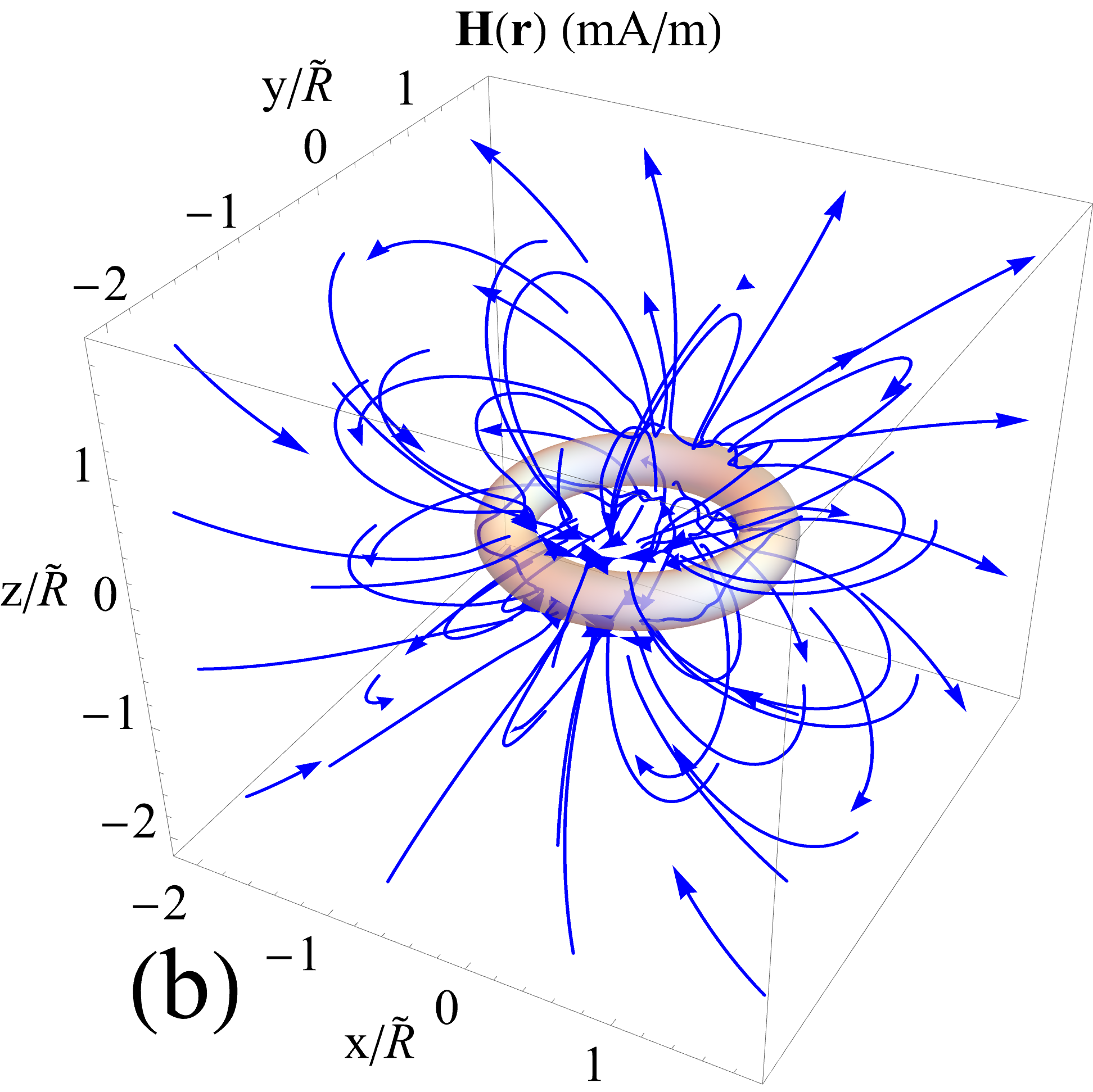}
\includegraphics[width=0.4\linewidth]{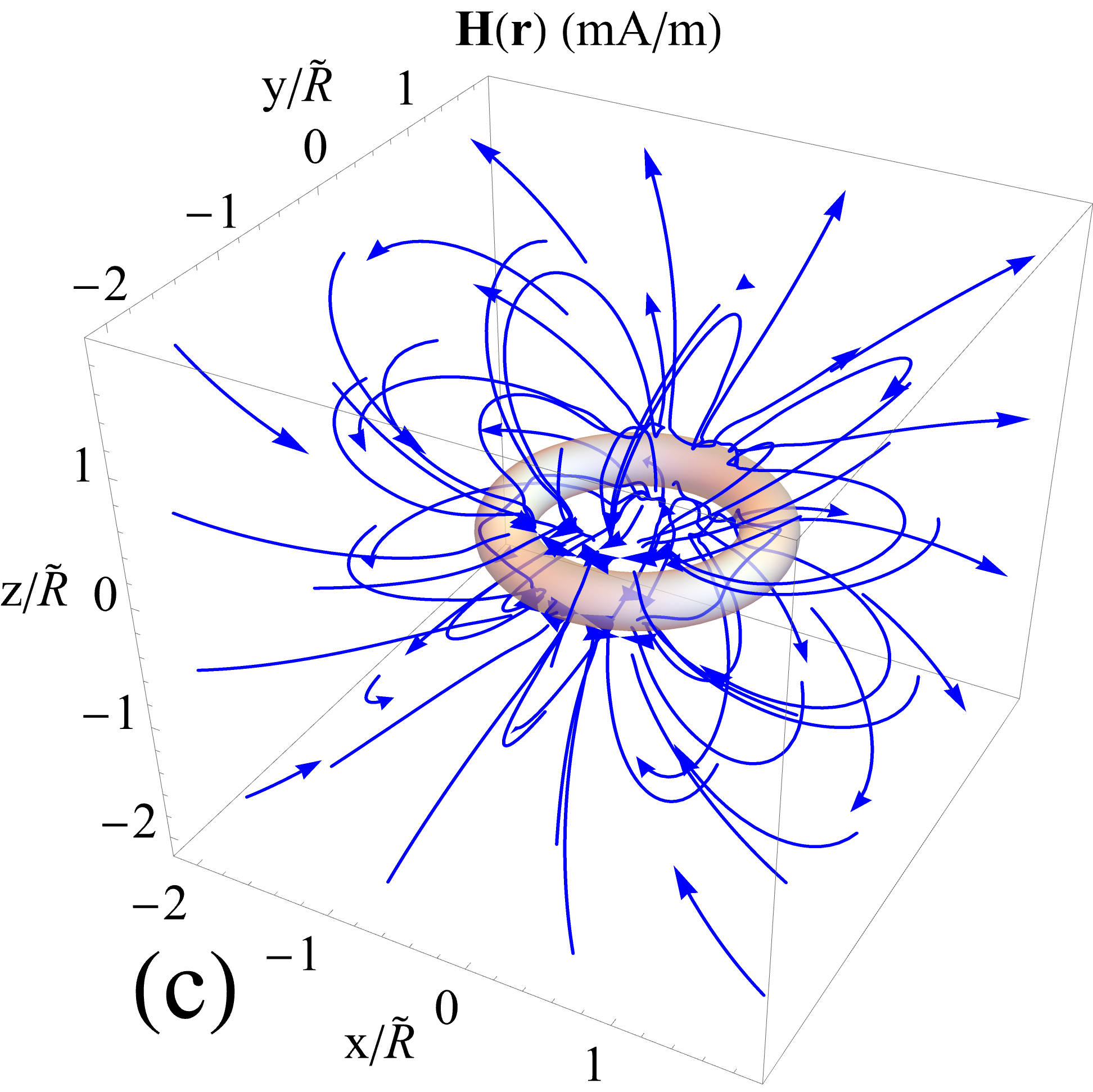}
\caption{As in Fig.~\ref{Fig5}, but for $\#m = \uy$.}
\label{Fig6}
\end{figure}

\begin{figure}[H]
 \centering
\includegraphics[width=0.4\linewidth]{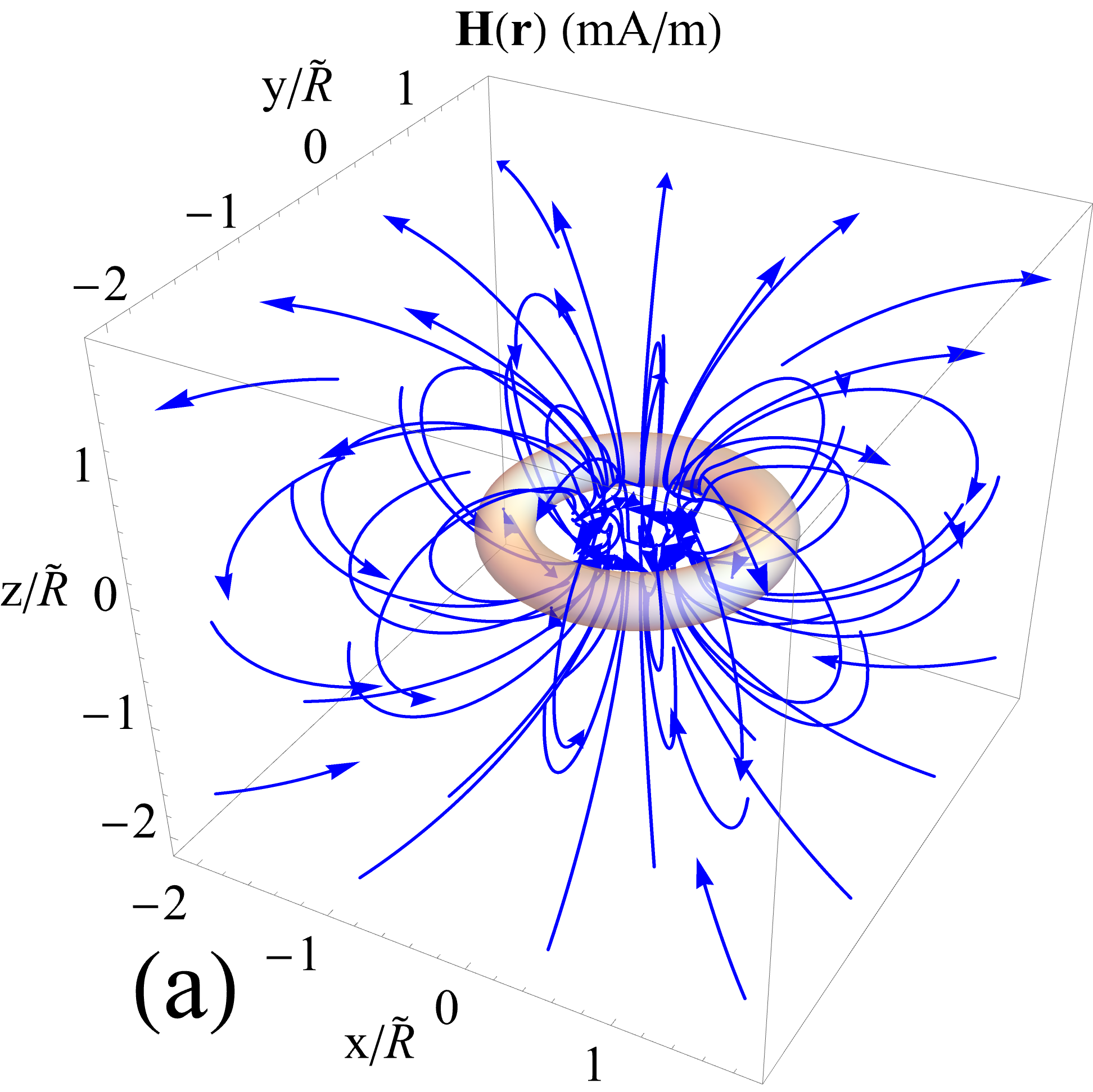}
\includegraphics[width=0.4\linewidth]{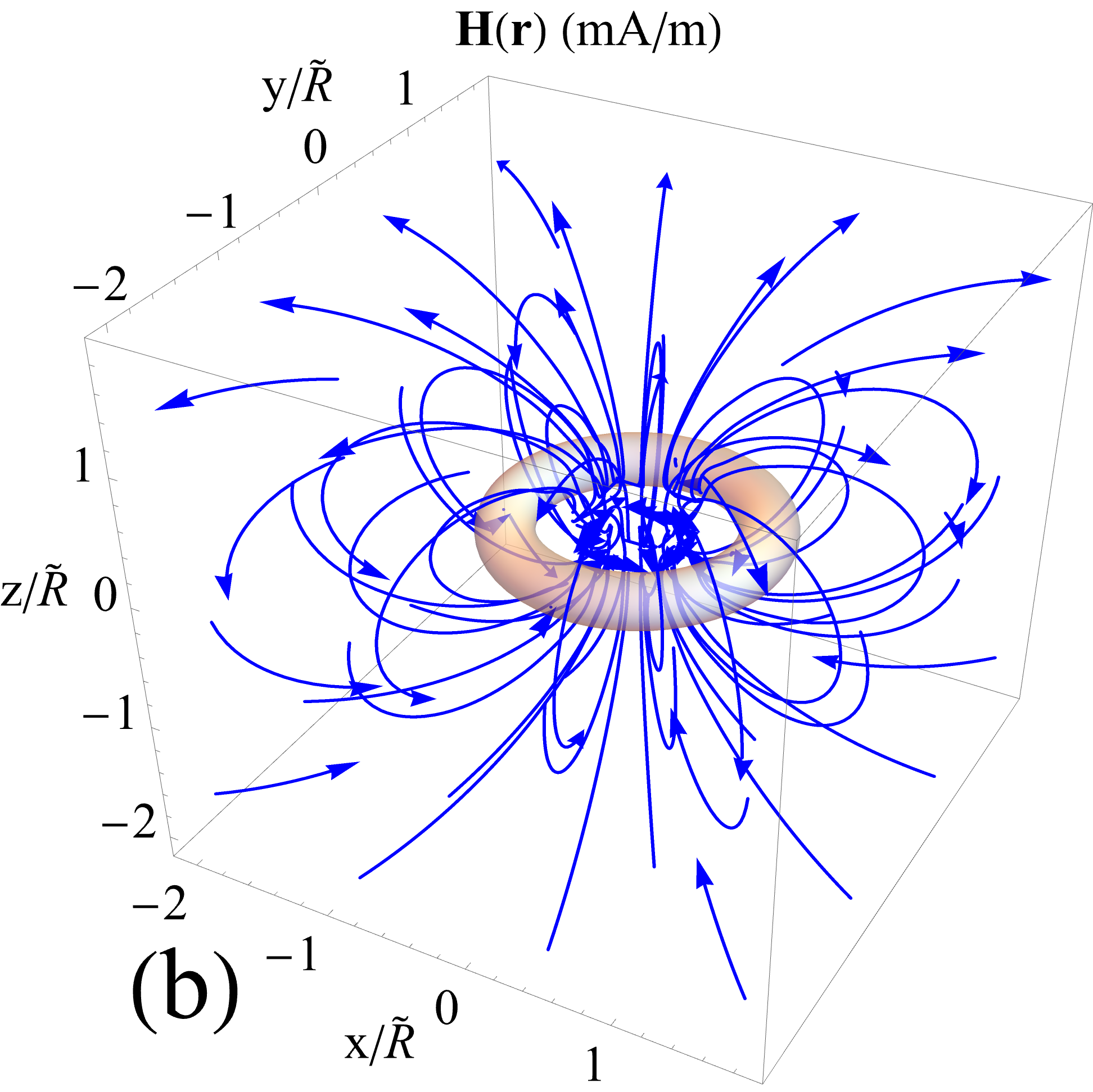}
\includegraphics[width=0.4\linewidth]{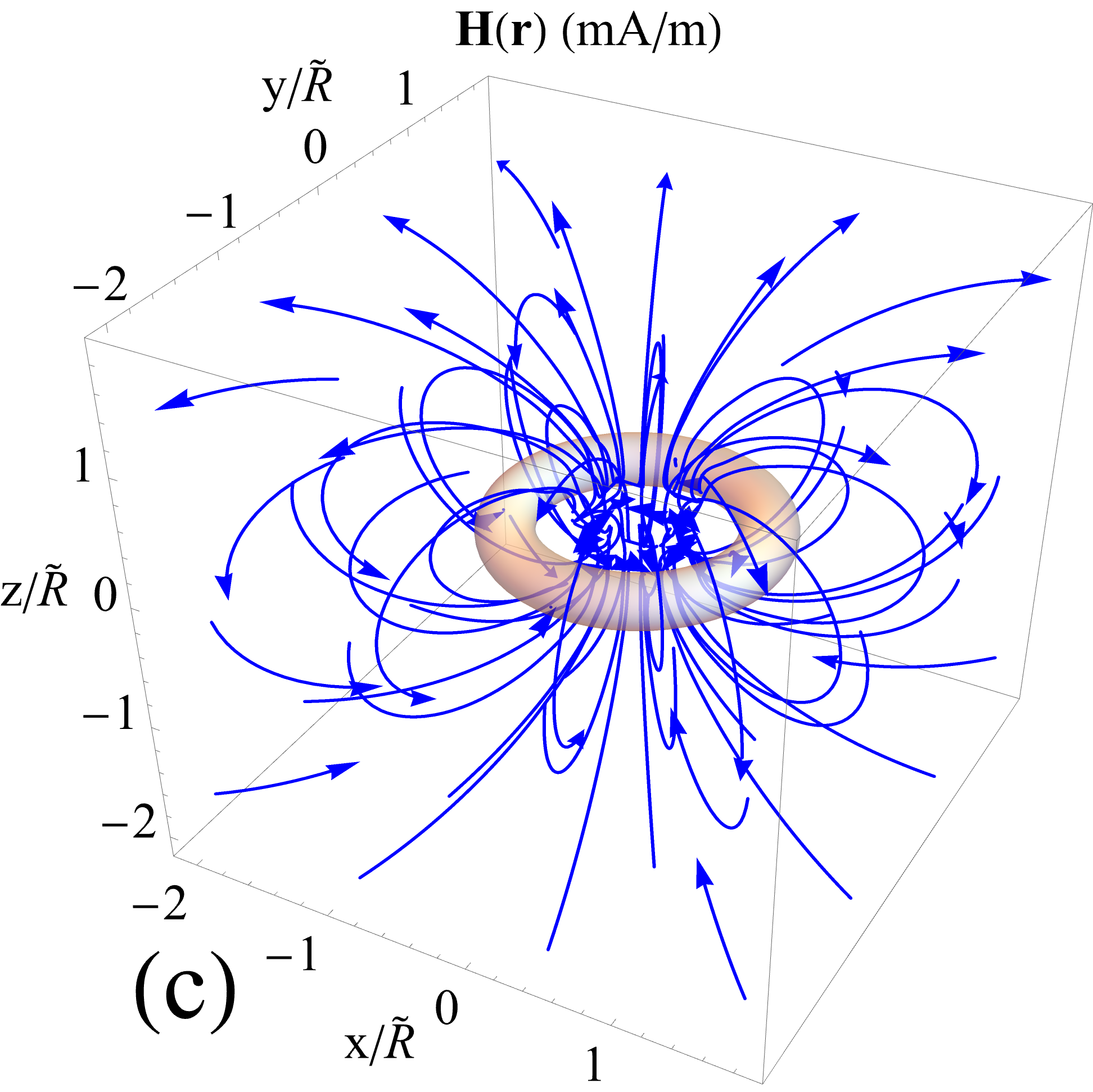}
\caption{As in Fig.~\ref{Fig5}, but for $\#m = \uz$.}
\label{Fig7}
\end{figure}

Now, we present the case of a point-magnetic-dipole source with the dipole located at the origin (i.e., $\xio = 0$, $\etao = \pi$).
Figures~\ref{Fig5}, \ref{Fig6}, and  \ref{Fig7} correspond to the cases $\#m = \ux$, $\#m = \uy$, and $\#m = \uz$, respectively.
In each of these three figures, vector plots of the total magnetostatic field as a function of $x \in [-2, 2] \tilde{R}$, $y \in [-2, 2] \tilde{R}$, and $z \in [-2, 2] \tilde{R}$ are presented, with the different panels corresponding to (a) $\alpha_x = 1$ and $\alpha_y = 1$, (b) $\alpha_x = 1.2$ and $\alpha_y = 1.2$, and (c) $\alpha_x = 1.1$ and $\alpha_y = 1.2$. In all figures, a closed-loop behavior of the total magnetostatic field lines is observed as is characteristic for a dipole excitation. In the regions near the toroid, the effect of anisotropy on the total magnetostatic field is evident in the sense that the total magnetostatic field changes substantially when anisotropy is introduced to the magnetic material of the toroid.

In general, by examining Figs.~\ref{Fig2}-\ref{Fig7}, it is observed that anisotropy of the toroid has a much stronger effect on the total magnetostatic field when the source is the field of a point magnetic dipole compared to when it is a uniform magnetostatic field.\\

\section{Concluding Remarks}\label{conc}
We have formulated a boundary-value problem to determine the perturbation of a magnetostatic field by a toroid made of a homogeneous anisotropic magnetic material. Whereas the solutions of the Laplace equation in the toroidal coordinate system were employed to express the total potential (and, therefore, the total magnetostatic field) as a series outside the toroid, a series of the toroidal solutions of the Laplace equation in a transformed space had to be used to express the  potential inside the toroid. Standard boundary conditions were then enforced on the surface of the toroid, and their orthogonality properties on the toroidal surface were used to determine a transition matrix. This transition matrix relates
the perturbation-potential coefficients to the source-potential coefficients.
Illustrative numerical results show that anisotropy of the toroid material has significant effects on the magnetostatic field in the region near the toroid.

\section*{Appendix}

The closed-form free-space Green function
\begin{equation} \label{Green-closed}
G(\#r, \bro) = \frac{1}{\vert\#r-\bro\vert},
\end{equation}
where $\#r\equiv(\xi,\eta,\phi)$ and $\bro\equiv(\xio,\etao,\phio)$,
can be expanded in bilinear form in toroidal coordinates as \cite{Bates}
\begin{equation} \label{Green-bilinear}
\begin{aligned}
G(\#r, \bro) &=  \sum_{j\in\left\{e,o\right\}}^{} \sum_{k\in\left\{e,o\right\}}^{} \sum_{\ell=0}^{\infty} \sum_{n=0}^{\infty}  \frac{\left(2-\delta_{{\ell}0}\right)\left(2-\delta_{n0} \right)(-1)^{\ell} \Gamma \left(n-{\ell}+\frac{1}{2} \right)}{\pi c \, \Gamma \left(n+{\ell}+\frac{1}{2} \right)} \\
& \times \left\{ \begin{array}{ll} \Psione(\#r)  \Psitwo(\bro) , & \xi > \xio,  \\ [5pt]
\Psione(\bro)  \Psitwo(\#r), & \xi < \xio.
\end{array} \right.
\end{aligned}
\end{equation}

The source potential due to a point dipole $\#m = {|\vec{m}|} \, \um$ located at $\bro\equiv(\xio,\etao,\phio)$, with $\xio < a$, is given by \cite{Zangwill}
\begin{equation}
\Phi\inc(\#r)=\frac{\muo {|\vec{m}|}}{4\pi}\,\um\.\nabla_{\rm o}\left(\frac{1}{\vert\#r-\bro\vert}\right)\,,
\end{equation}
where  $\nabla_{\rm o}(...)$ denotes the gradient with respect to $\bro$. Equation (\ref{Green-bilinear})
can be used to expand the source potential as
\begin{equation}
\Phi\inc(\#r)= \sum \limits_{j\in\left\{e,o\right\}}^{} \sum \limits_{k\in\left\{e,o\right\}}^{}\sum\limits_{{\ell}=0}^{\infty} \sum\limits_{n=0}^{\infty} \left\{ \begin{array}{ll}
\calA\jkmn \Psione(\#r), & \xi > \xio,       \\[5pt]
\bar{\calA}\jkmn \Psitwo(\#r), & \xi < \xio,
\end{array} \right.
\end{equation}
where the coefficients
\begin{subequations}
\begin{equation}
\label{A-ps-p}
\calA\jkmn=  \frac{ \muo {|\vec{m}|}}{4 \pi^2 c} \frac{\left(2-\delta_{{\ell}0}\right)\left(2-\delta_{n0} \right)(-1)^{\ell} \Gamma \left(n-{\ell}+\frac{1}{2} \right)}{\Gamma \left(n+{\ell}+\frac{1}{2} \right)} \, \um \. \nabla_{\rm o}\left[\Psitwo(\bro) \right]
\end{equation}
and
\begin{equation}
\label{Atilde-ps-p}
\bar{\calA}\jkmn = \frac{ \muo {|\vec{m}|}}{4 \pi^2 c} \frac{\left(2-\delta_{{\ell}0}\right)\left(2-\delta_{n0} \right)(-1)^{\ell} \Gamma \left(n-{\ell}+\frac{1}{2} \right)}{\Gamma \left(n+{\ell}+\frac{1}{2} \right)} \, \um \. \nabla_{\rm o}\left[\Psione(\bro) \right].
\end{equation}
\end{subequations}

The bilinear form of the Green function presented in Eq. (\ref{Green-bilinear}) was originally   derived by Morse and Feshbach \cite{Morse}. However, as pointed by Bates  \cite{Bates}, their expression does not contain the constant $c$ {and} involves typographical errors. After evaluating the source potential due to a point magnetic dipole on the interface $\xi = a$ using (i) Eq. (\ref{Green-closed}), (ii) Eq. (\ref{Green-bilinear}), and (iii) the bilinear form of the Green function reported by Morse and Feshbach \cite{Morse}, calculations corresponding to various dipole orientations $\um$ (not shown) have confirmed that Eq. (\ref{Green-bilinear}) is the correct bilinear form of Eq. (\ref{Green-closed}).

\vspace{5mm}
\noindent\textbf{Acknowledgement.} AL is grateful to the Charles Godfrey Binder Endowment at Penn State for supporting
his research activities from 2006 to 2024. H.~M.
Alkhoori acknowledges the United Arab Emirates University for supporting this work.

\vspace{5mm}


\begin{thebibliography}{99}

\bibitem{Eisenhart}
L. P. Eisenhart,  Separable systems in euclidean 3-space, \textit{Physical Review} \textbf{45}, (1934) 427--428.

\bibitem{Moon}
P. Moon and D. E. Spencer, \textit{Field Theory Handbook: Including Coordinate Systems, Differential Equations and Their Solutions}, 2nd ed., (Berlin, Germany, Springer 1971).

\bibitem{Chen}
H. C. Chen, \textit{Theory of Electromagnetic Waves: A Coordinate-Free Approach}
(New York, USA, McGraw--Hill 1983).

\bibitem{EAB}
T. G. Mackay and A. Lakhtakia, \textit{Electromagnetic Anisotropy and Bianisotropy}, 2nd ed.
(Singapore, World Scientific 2020).

\bibitem{McCaig}
M. McCaig, \textit{Permanent Magnets in Theory and Practice} (London, United Kingdom, Pentech 1977).

\bibitem{Gersten}
J. I. Gersten and F. W. Smith, \textit{The Physics and Chemistry of Materials} (New York, USA, Wiley 2001).



\bibitem{Meiklejohn}
W. H. Meiklejohn and C. P. Bean,
New magnetic anisotropy, \textit{Physical Review} \textbf{105} (1957) 904--913.

\bibitem{Johnson}
M. T. Johnson, P. J. H. Bloemen, F. J. A. den~Broeder, and J. J. de~Vries,
Magnetic anisotropy in metallic multilayers, \textit{Reports of Progress in Physics} \textbf{59} (1996) 1409--1458.



\bibitem{Morse}
P. M. Morse and H. Feshbach,
\textit{Methods of Theoretical Physics}, Vol. II
(New York, USA, McGraw--Hill 1953).

\bibitem{Abramowitz}
M. Abramowitz and I. A. Stegun,
\textit{Handbook of Mathematical Functions} (New York, USA,
Dover Publications 1972).

\bibitem{Bates}
J. W. Bates,
On toroidal Green’s functions,
\textit{Journal of Mathematical Physics} \textbf{38} (1997) 3679--3691.

\bibitem{QJMAM1}
A. Lakhtakia, N. L. Tsitsas, and H. M. Alkhoori, Theory of perturbation of
electrostatic field by an anisotropic sphere,
\textit{Quarterly Journal of Mechanics and Applied Mathematics} \textbf{74} (2021) 467--490;
correction: \textbf{75} (2022) 125.



\bibitem{Zangwill}
A. Zangwill,
\textit{Modern Electrodynamics}
(New York, USA, Cambridge University Press 2013).








\end{thebibliography}
\end{document}